\documentclass[aps,prd,twocolumn,amsmath,amsfonts,amssymb,nofootinbib,eqsecnum,secnumarabic,notitlepage,superscriptaddress,10pt]{revtex4-1}
\usepackage[dvipsnames]{xcolor}
\usepackage{graphicx,hyperref}
\hypersetup{colorlinks=true,linkcolor=ForestGreen,citecolor=ForestGreen,filecolor=ForestGreen,urlcolor=ForestGreen}
\DeclareUnicodeCharacter{03BC}{\ensuremath{\mu}}

\begin{document}
\title{Multiboson Signatures of Doubly Charged Scalars at a Same-Sign Muon Collider}

\author{Mariana Frank}
\email{mariana.frank@concordia.ca}
\affiliation{Department of Physics, Concordia University, 7141 Sherbrooke St. West, Montreal, Quebec H4B 1R6, Canada}

\author{Benjamin Fuks}
\email{fuks@lpthe.jussieu.fr}
\affiliation{Laboratoire de Physique Théorique et Hautes Énergies (LPTHE), UMR 7589, Sorbonne Université et CNRS, 4 place Jussieu, 75252 Paris Cedex 05, France}

\author{Chayan Majumdar}
\email{chayanmajumdar@impcas.ac.cn}
\affiliation{Institute of Modern Physics, Chinese Academy of Sciences, Lanzhou, 730000, China}

\author{Supriya Senapati}
\email{ssenapati@njust.ac.cn}
\affiliation{Department of Applied Physics and MIIT Key Laboratory of Semiconductor Microstructure and Quantum Sensing, Nanjing University of Science and Technology, Nanjing 210094, China}

\begin{abstract}
  We investigate the sensitivity of a same-sign $\mu^+\mu^+$ collider to doubly charged scalars in the Type-II seesaw framework, focusing on the regime in which the doubly charged scalar decays dominantly into same-sign $W$-boson pairs. Motivated by the $\mu$TRISTAN proposal, for a benchmark we consider a degenerate triplet spectrum at a center-of-mass energy of $2~{\rm TeV}$ and an integrated luminosity of $1~{\rm ab}^{-1}$. The signal process is studied in the fully hadronic $W$-decay mode, leading to a characteristic $\mu^+\mu^+ + 8j$ final state. We develop a cut-based analysis based on high jet multiplicity and global hadronic $W$ reconstruction, and then improve its sensitivity with a multivariate strategy exploiting reconstructed $W$ observables, global event kinematics, multi-boson variables and spectator-muon information. The best-performing setup reaches a $2\sigma$ sensitivity level for doubly charged scalar masses up to $425-430~{\rm GeV}$, extending the cut-based reach by a few tens of GeV. This  indicates an improvement over the current LHC coverage, while also providing an independent probe based on a qualitatively different production mode and collider environment.

\end{abstract}

\pacs{}
\maketitle

\section{Introduction} \label{sec:intro}

Doubly charged scalars are predicted in a variety of extensions of the Standard Model (SM), and they constitute one of the most striking signatures of extended scalar sectors. A particularly well-motivated example is provided by the Type-II seesaw mechanism, in which a complex ${\rm SU}(2)_L$ scalar triplet is introduced to generate Majorana neutrino masses~\cite{Konetschny:1977bn, Cheng:1980qt, Magg:1980ut, Lazarides:1980nt, Schechter:1980gr, Mohapatra:1980yp}, so that for a hypercharge-one triplet, the scalar spectrum necessarily contains a doubly charged component. The structure of the scalar potential, the resulting mass spectrum and the associated theoretical consistency constraints have been analyzed in detail in Refs.~\cite{Arhrib:2011uy, Bonilla:2015eha, Cai:2017mow}, and the collider phenomenology of the doubly charged state is closely tied to the neutrino-mass-generation mechanism and to the value of the triplet vacuum expectation value~\cite{Muhlleitner:2003me, Akeroyd:2005gt, Han:2007bk, Mitra:2016wpr, Fuks:2019clu}. 

Doubly charged scalars also arise in other well-known frameworks. In left-right symmetric models, including supersymmetric realizations, they appear as components of scalar triplets associated with the breaking of the extended gauge symmetry~\cite{Pati:1974yy, Mohapatra:1974gc, Mohapatra:1974hk, Senjanovic:1975rk, Mohapatra:1979ia}, and their charged-scalar phenomenology has been studied in several collider and low-energy contexts~\cite{Deshpande:1990ip, Dutta:2014dba, Basso:2015pka, BhupalDev:2016nfr, Borah:2018yxd, BhupalDev:2018tox}. They are also present in radiative neutrino-mass models such as the Zee-Babu model~\cite{Zee:1985id, Babu:1988ki, Nebot:2007bc, Herrero-Garcia:2014hfa, Schmidt:2014zoa}, in custodial-symmetric triplet extensions such as the Georgi-Machacek model~\cite{Georgi:1985nv, Chanowitz:1985ug} whose doubly charged fiveplet state has been studied in collider and global-fit analyses~\cite{Sun:2017mue, Chiang:2018cgb}, and in other scalar-sector extensions involving, for instance, hypercharge-$3/2$ doublets and ${\rm SU}(2)_L$-singlet doubly charged scalars~\cite{Aoki:2011pz, Crivellin:2018ahj}, as well as non-minimal triplet or more general exotic scalar sectors~\cite{Gunion:1989ci, Chaudhuri:2013xoa, Chala:2018opy}.

The collider phenomenology associated with doubly charged scalars has consequently been studied in several complementary directions. Early and systematic hadron-collider analyses focused mainly on Drell-Yan pair production, associated production and multilepton signatures in triplet-scalar scenarios~\cite{Huitu:1996su, Akeroyd:2005gt, Han:2007bk, FileviezPerez:2008wbg, FileviezPerez:2008jbu, Akeroyd:2010ip, Alloul:2013raa, Mitra:2016wpr, Frank:2025jjt}. More general treatments have considered lepton-number-violating signatures mediated by singly and doubly charged scalars, singlet doubly charged states, and higher-order or future-collider aspects of doubly charged scalar production~\cite{delAguila:2013mia, King:2014uha, Fuks:2019clu, Guedes:2025jqu}. Furthermore, in regimes where the diboson decay mode becomes important, dedicated studies have investigated same-sign $W$-boson final states at the LHC~\cite{Chiang:2012dk, Kanemura:2013vxa, Kang:2014lwn, Kanemura:2014ipa}, and vector-boson-fusion production mechanisms have been studied at the LHC and future hadron colliders~\cite{Huitu:1996su, Dutta:2014dba, Bambhaniya:2015wna}. Lepton-colliders then additionally provide a complementary environment, especially for same-sign muon-collider configurations~\cite{Das:2024kyk, Chiang:2025lab}.

In Type-II seesaw models, the existing bounds on the doubly charged scalar bosons $\Delta^{\pm\pm}$ are, however, highly sensitive to the region of the parameter space under consideration. The relevant phenomenology is controlled primarily by the triplet vacuum expectation value (vev) $v_t$, by the mass of the doubly charged scalar, and by the mass splittings among the charged and neutral triplet components~\cite{Melfo:2011nx, Primulando:2019evb}. In particular, the value of $v_t$ plays a central role in determining the dominant decay modes of the doubly charged scalar. 

For sufficiently small values, typically $v_t\lesssim 10^{-4}~{\rm GeV}$, the Yukawa-induced dileptonic modes $\Delta^{\pm\pm}\to \ell^\pm\ell'^\pm$ dominate. Under the assumption of prompt leptonic decays and equal branching fractions into the six same-sign dilepton flavor combinations, the ATLAS collaboration constrained this setup and obtained a lower bound of $m_{\Delta^{\pm\pm}}>1080~{\rm GeV}$ at $95\%$ confidence level~\cite{ATLAS:2022pbd}. This limit is therefore representative of the leptonic regime, but it is not model-independent and depends on the assumed flavor pattern of the triplet Yukawa couplings. 

For larger values of $v_t$, still compatible with electroweak precision tests and extending up to at most a few GeV, the dileptonic branching fractions are suppressed, while the gauge-boson mode $\Delta^{\pm\pm}\to W^\pm W^\pm$ becomes dominant when cascade decays into lighter triplet states are absent or subleading. In the earlier ATLAS search of Ref.~\cite{ATLAS:2018ceg}, doubly charged scalar masses in the range $200-220~{\rm GeV}$ are excluded, while for the more recent analysis of Ref.~\cite{ATLAS:2021jol}, two benchmark scenarios are considered. The `pair-production' benchmark assumes a singly charged scalar to be at least $100~{\rm GeV}$ heavier than the doubly charged state, whereas the `associated-production' benchmark considers nearly degenerate singly and doubly charged scalars with a mass splitting below $5~{\rm GeV}$. The corresponding lower limits on the doubly charged scalar mass reach $350~{\rm GeV}$ and $230~{\rm GeV}$, respectively, with the former applying to scenarios well described by the pair-production-only
benchmark. More generally, electroweak precision constraints restrict, but do not eliminate, the possibility of sizable mass splittings among the triplet components, and mass splittings of order several tens of GeV can remain viable for masses at the few-hundred-GeV scale~\cite{Melfo:2011nx, Chun:2012jw}. For non-degenerate spectra, scalar-to-scalar cascade decays can then compete with, or even dominate over, the direct dileptonic and diboson modes in sizable regions of the parameter space~\cite{Melfo:2011nx, Primulando:2019evb}. The collider phenomenology of the triplet scalars is therefore qualitatively different in the leptonic, diboson, and cascade-dominated regimes.

The proposed $\mu$TRISTAN program provides a particularly interesting environment in which to revisit such questions. Based on the acceleration of ultra-cold antimuons developed in the context of the J-PARC muon program, $\mu$TRISTAN would allow both a $\mu^+e^-$ collider mode at $\sqrt{s}=346~{\rm GeV}$ and a same-sign $\mu^+\mu^+$ mode reaching $\sqrt{s}=2~{\rm TeV}$~\cite{Abe:2019thb, Hamada:2022mua, Hamada:2022uyn}. Its clean leptonic environment, same-sign initial state and TeV-scale energy reach make it especially well suited to testing new interactions connected with lepton flavor, lepton number and the origin of neutrino masses. In particular, $\mu$TRISTAN has been shown to be sensitive to the signatures of leptophilic neutral and doubly charged scalars such as those arising in neutrino-mass models~\cite{Fridell:2023gjx, Dev:2023nha, Lichtenstein:2023iut}, and same-sign muon collisions have additionally been explored as probes of lepton-flavor and lepton-number violation, heavy Majorana neutrinos, heavy neutral leptons and multi-lepton signatures in a variety of beyond the SM settings~\cite{Jiang:2023mte, Li:2023lkl, deLima:2024ohf, Bhattacharya:2025xwv, Dehghani:2025xkd, Kitano:2025xaj, Yan:2026wrb}. These studies illustrate the broader potential of $\mu$TRISTAN and motivate a dedicated analysis of the Type-II seesaw scalar sector in the poorly constrained diboson-dominated regime.

In the present work, we focus on the large-$v_t$ regime and choose a degenerate triplet spectrum so that cascade decays are absent and the doubly charged scalar decays dominantly into same-sign $W$-boson pairs. We investigate the sensitivity of the $\sqrt{s}=2~{\rm TeV}$ same-sign $\mu^+\mu^+$ configuration to doubly charged scalar production and decay in the Type-II seesaw model, assuming an integrated luminosity of $1~{\rm ab}^{-1}$. In this setup, doubly charged scalar-pair production proceeds through electroweak vector-boson-fusion topologies, $\mu^+\mu^+\to\mu^+\mu^+\Delta^{++}\Delta^{--}$, leaving two spectator muons in the final state. Taking all four $W$ bosons to decay hadronically then maximizes the signal branching fraction and leads to a characteristic $\mu^+\mu^+ + 8j$ final state, at the cost of a challenging combinatorial reconstruction.

Our work is organized as follows. In Section~\ref{sec:model}, we introduce the model setup, the relevant decay modes of the doubly charged scalar and briefly detail the current collider limits. In Section~\ref{sec:analysis}, we first develop a cut-based analysis exploiting the high jet multiplicity of the considered signature and the reconstruction of multiple hadronic $W$ candidates, and then improve the sensitivity through a boosted-decision-tree multivariate analysis based on reconstructed-$W$ observables, global event kinematics, multi-boson variables and spectator-muon information. We summarize our results and comment on the outlook in Section~\ref{sec:summary}.

\section{Doubly Charged Scalars in the Type-II Seesaw Model}
\label{sec:model}

\subsection{Theoretical considerations and constraints}
\label{subsec:theory}

We consider the Type-II seesaw framework, which provides a minimal realization of Majorana neutrino mass generation through the addition of a scalar $SU(2)_L$ triplet whose neutral component acquires a naturally small vacuum expectation value~\cite{Magg:1980ut, Lazarides:1980nt, Cheng:1980qt, Mohapatra:1980yp, Schechter:1980gr}. In this framework, the scalar sector of the SM is extended by a complex triplet field transforming as $\Delta \sim (1,3,2)$ under the SM gauge group $SU(3)_c\times SU(2)_L\times U(1)_Y$, where we use the convention $Q=T_3+Y/2$. In matrix form, the triplet can be written as
\begin{equation}\renewcommand{\arraystretch}{1.2}
\Delta=
\begin{pmatrix}
\frac{\delta^+}{\sqrt{2}} & \delta^{++} \\
\delta^{0} & -\frac{\delta^+}{\sqrt{2}}
\end{pmatrix}.
\end{equation}
Together with the SM Higgs doublet $H=(\phi^+,\phi^0)^T$, the full scalar-sector Lagrangian reads
\begin{equation}
  \mathcal{L} \supset (D_\mu H)^\dagger(D^\mu H) + \mathrm{Tr} \left[(D_\mu \Delta)^\dagger (D^\mu \Delta) \right] - V(H,\Delta),
\end{equation}
where the most general renormalizable scalar potential is given by~\cite{Arhrib:2011uy, Bonilla:2015eha, Fuks:2019clu}
\begin{equation}\begin{split}
  & V(H,\Delta) = - m_H^2 H^\dagger H + m_\Delta^2 \rm{Tr} (\Delta^\dagger \Delta)  \\[.1cm]
    &\quad  + \left[ \mu H^T i \sigma_2 \Delta^\dagger H +\rm{H.c.} \right] + \lambda_H (H^\dagger H)^2 \\[.1cm]
    &\quad + \lambda_1 (H^\dagger H) \rm{Tr} (\Delta^\dagger \Delta) + \lambda_2 \left[ \rm{Tr} (\Delta^\dagger \Delta) \right]^2 \\[.1cm]
    &\quad + \lambda_3 \mathrm{Tr} \left[ (\Delta^\dagger \Delta)^2 \right] + \lambda_4 H^\dagger \Delta \Delta^\dagger H.
\end{split}\label{eq:Potential}\end{equation}
Here $m_H$ and $m_\Delta$ denote two mass parameters with $m_H^2>0$, while $\lambda_H$ and $\lambda_i$ correspond to the different quartic couplings. Assigning a lepton number $L(\Delta)=-2$ to the triplet field, the triplet Yukawa interaction with the weak left-handed SM lepton doublet, 
\begin{equation}
  \mathcal{L}_{Y_\Delta} = - Y_{\Delta, ij} L_i^{T} C i \sigma_2 \Delta L_j + \rm{H.c.},
\label{eq:Lyuk}\end{equation}
with $i,j$ being flavor indices and $Y_\Delta$ being a complex symmetric matrix, conserves lepton number. In contrast, the trilinear $\mu$-term of the potential explicitly violates it by two units, such that the limit $\mu\to 0$ restores lepton number conservation, rendering a small value of $\mu$ technically natural. After electroweak symmetry breaking, we express the neutral scalar components as $\phi^0=(v_d+h+iG^0)/\sqrt{2}$ and $\delta^0=(v_t+\delta_R+i\delta_I)/\sqrt{2}$, which then acquire the vevs
\begin{equation}
  \langle H \rangle = \frac{1}{\sqrt{2}} \begin{pmatrix} 0\\ v_d \end{pmatrix}, \qquad
  \langle \Delta \rangle = \frac{1}{\sqrt{2}} \begin{pmatrix} 0 & 0 \\ v_t & 0 \end{pmatrix}.
\end{equation}

With these conventions, the minimization conditions of the scalar potential read
\begin{equation}\begin{split}
m_H^2 &= \lambda_H v_d^2 + \frac{v_t^2}{2}(\lambda_1+\lambda_4) - \sqrt{2}\mu v_t , \\
m_\Delta^2 &= \frac{\mu v_d^2}{\sqrt{2}v_t} - \frac{v_d^2}{2}(\lambda_1+\lambda_4) - v_t^2(\lambda_2+\lambda_3).
\end{split}\label{eq:min}\end{equation}
For $v_t\ll v_d$ and neglecting terms of order $v_t^3$, the second condition implies approximately
\begin{equation}
  v_t \simeq \frac{\mu\, v_d^2} {\sqrt{2}\left[m_\Delta^2+\frac{1}{2}(\lambda_1+\lambda_4)v_d^2\right]},
\end{equation}
which shows explicitly that the triplet vev is induced by the lepton-number-violating parameter $\mu$ and vanishes in the lepton-number-conserving limit $\mu\to0$. Equivalently, for a fixed $\mu$ value, a small $v_t$ vev can be obtained from a large effective triplet mass scale,
\begin{equation}
  M_\Delta^2 \equiv m_\Delta^2+\frac{1}{2}(\lambda_1+\lambda_4)v_d^2,
\end{equation}
which illustrates the seesaw suppression characteristic of the Type-II mechanism. In addition, the triplet vev contributes to the electroweak gauge-boson masses according to
\begin{equation}\begin{split}
  m_W^2 =&\ \frac{g^2}{4}\left(v_d^2+2v_t^2\right), \\
  m_Z^2 =&\ \frac{g^2+g^{\prime 2}}{4} \left(v_d^2+4v_t^2\right),
\end{split}\end{equation}
so that the electroweak scale is
\begin{equation}
v^2 \equiv v_d^2+2v_t^2 \simeq (246~{\rm GeV})^2 .
\end{equation}
The tree-level value of the electroweak $\rho$ parameter is therefore modified as
\begin{equation}
  \rho \equiv \frac{m_W^2}{m_Z^2\cos^2\theta_W} = \frac{v_d^2+2v_t^2}{v_d^2+4v_t^2}.
\end{equation}
For nonzero $v_t$, this expression is smaller than unity and therefore corresponds to a negative tree-level shift with respect to the SM prediction. Electroweak precision measurements consequently impose an upper bound on the triplet vev of order a few GeV~\cite{Kanemura:2012rs, Cheng:2022jyi}, often equivalently discussed in terms of the custodial symmetry breaking contribution to the oblique $T$ parameter~\cite{Peskin:1991sw}. This deviation reflects the breaking of the custodial-symmetry relation by the $Y=2$ triplet vev, in contrast with the SM doublet contribution which preserves $\rho=1$ at tree level. In the following, we adopt the representative constraint $v_t\lesssim 4.8~{\rm GeV}$ at $95\%$ confidence level~\cite{Primulando:2019evb}.

After electroweak symmetry breaking, the Yukawa interaction in Eq.~\eqref{eq:Lyuk} generates the Majorana neutrino mass matrix
\begin{equation}
  M_\nu = \sqrt{2}\,Y_\Delta v_t,
\end{equation}
which is diagonalized by the Pontecorvo-Maki-Nakagawa-Sakata (PMNS) matrix according to
\begin{equation}
  M_\nu = U_{\rm PMNS}^\ast\, m_\nu^{\rm diag}\, U_{\rm PMNS}^\dagger,
\end{equation}
where $m_\nu^{\rm diag}$ contains the three light-neutrino masses. The triplet Yukawa matrix is therefore given by
\begin{equation}
Y_{\Delta,ij} = \frac{(M_\nu)_{ij}}{\sqrt{2}v_t},
\end{equation}
such that its flavor structure is fixed by neutrino oscillation data up to the choice of the neutrino mass ordering, the lightest neutrino mass and the different Majorana phases which remain largely unconstrained~\cite{Esteban:2024eli}. Consequently, the relative branching fractions related to the decays $\Delta^{\pm\pm} \to \ell_i^\pm\ell_j^\pm$ are controlled by the entries of the neutrino mass matrix~\cite{Garayoa:2007fw, Kadastik:2007yd}, as will be further discussed below. Since $Y_\Delta\propto 1/v_t$ for fixed neutrino masses, low-energy charged-lepton-flavor-violation (CLFV) observables, such as $\mu\to e\gamma$, $\mu\to 3e$ and $\tau\to\ell\ell\ell$, constrain combinations of $Y_{\Delta,ij}$ and the triplet-scalar masses. These constraints can be important in the small-$v_t$ regime~\cite{Chun:2003ej, Dev:2018sel}. However, for large triplet vevs $v_t\sim{\cal O}(1)$ GeV, the Yukawa couplings required to reproduce the observed neutrino masses are extremely small, $Y_\Delta\sim 10^{-11}$ for a representative neutrino mass scale of $0.05$ eV, so these CLFV constraints are strongly suppressed. 

After electroweak symmetry breaking, the scalar spectrum contains the SM-like Higgs boson $h$, a doubly charged scalar pair $\Delta^{\pm\pm}$, a singly charged scalar pair $\Delta^\pm$, and two additional neutral states conventionally denoted by the CP-even scalar $\Delta^0$ and the CP-odd scalar $\chi$. The remaining three scalar degrees of freedom correspond to the Goldstone bosons absorbed by the $W^\pm$ and $Z$ bosons. For $v_t\ll v_d$, the charged and neutral Goldstone modes are mostly doublet-like, while the physical triplet-like states mix with the doublet fields only through angles suppressed by $v_t/v_d$. After using the minimization conditions to trade the two mass parameters of the potential $m_H^2$ and $m_\Delta^2$ for the vevs, and expressing $\mu$ in terms of $v_t$, the scalar sector can be conveniently parameterized by seven independent inputs consisting of the triplet vev $v_t$, two quartic couplings $\lambda_{1,2}$ and four physical scalar masses $m_h$, $m_{\Delta^{\pm\pm}}$, $m_{\Delta^\pm}$ and $m_{\Delta^0}$, 
\begin{equation}
\lambda_1,\quad
\lambda_2,\quad
m_h,\quad
m_{\Delta^{\pm\pm}},\quad
m_{\Delta^\pm},\quad
m_{\Delta^0},\quad
v_t .
\end{equation}
The remaining parameters of the potential, namely $\lambda_H$, $\lambda_3$, $\lambda_4$ and $\mu$, are then fixed by these inputs through the relations given in Ref.~\cite{Fuks:2019clu}. The benchmark choices used in this work are further required to yield a bounded-from-below scalar potential and to satisfy perturbative unitarity, perturbativity and electroweak precision constraints~\cite{Arhrib:2011uy, Aoki:2012jj, Primulando:2019evb}.

In the limit $v_t\ll v_d$, the scalar masses are approximately~\cite{FileviezPerez:2008jbu, Melfo:2011nx, Fuks:2019clu}
\begin{equation}\begin{split}
  m^2_{\Delta^{\pm\pm}} \simeq m_\Delta^2 - \frac{\lambda_4}{2}v_d^2,\qquad
  &m^2_{\Delta^\pm} \simeq m_\Delta^2 - \frac{\lambda_4}{4}v_d^2,\\
  m_h^2 \simeq 2\lambda_H v_d^2,\qquad
  & m^2_{\Delta^0} \simeq m_\chi^2 \simeq m_\Delta^2.
\end{split}\end{equation}
The degeneracy between the CP-even and CP-odd neutral triplet-like states is lifted only by terms suppressed by the triplet vev, with $m_{\Delta^0}^2-m_\chi^2={\cal O}(\mu v_t)$, and is therefore numerically negligible for $v_t\ll v_d$. In addition, the $\Delta$ mass-squared splittings obey the relation
\begin{equation}
  \Delta m^2 \equiv m^2_{\Delta^0} - m^2_{\Delta^\pm} \simeq m^2_{\Delta^\pm} - m^2_{\Delta^{\pm\pm}} \simeq \frac{\lambda_4}{4}v_d^2.
\end{equation}
At leading order in $v_t/v_d$, this equal spacing in mass squared is thus controlled entirely by the quartic interaction $\lambda_4 H^\dagger\Delta\Delta^\dagger H$ which splits the different electroweak components of the triplet after electroweak symmetry breaking. On the other hand, the remaining quartic couplings shift the triplet masses without splitting the electroweak components at this order and in the limit $v_t\to 0$. Thus, while $v_t$ controls the tree-level shift in the $\rho$ parameter, the $\lambda_4$ quartic coupling controls the custodial-sensitive mass splittings that enter electroweak precision observables at loop level~\cite{Blank:1997qa, Cheng:2022jyi}. For $\lambda_4=0$, the triplet spectrum is nearly degenerate at tree level, whereas for $\lambda_4>0$, one obtains $m_{\Delta^{\pm\pm}} < m_{\Delta^\pm} < m_{\Delta^0,\chi}$. By contrast, for $\lambda_4<0$ the hierarchy is reversed, $m_{\Delta^{\pm\pm}} > m_{\Delta^\pm} > m_{\Delta^0,\chi}$.  Throughout this work, we adopt the nearly degenerate scenario, corresponding to $\lambda_4\simeq0$.

\begin{figure*}
    \includegraphics[width=0.49\textwidth]{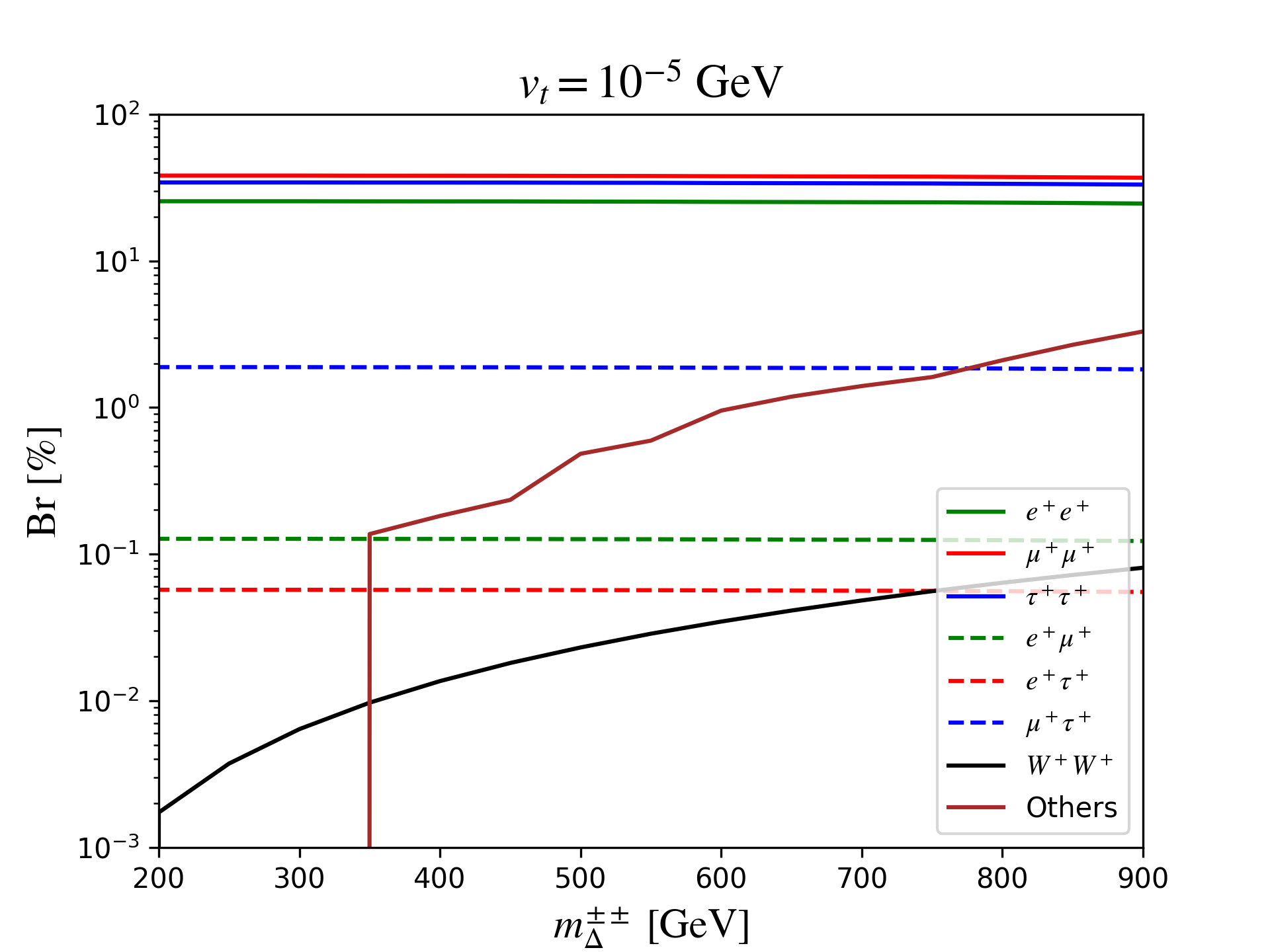}\hfill
    \includegraphics[width=0.49\textwidth]{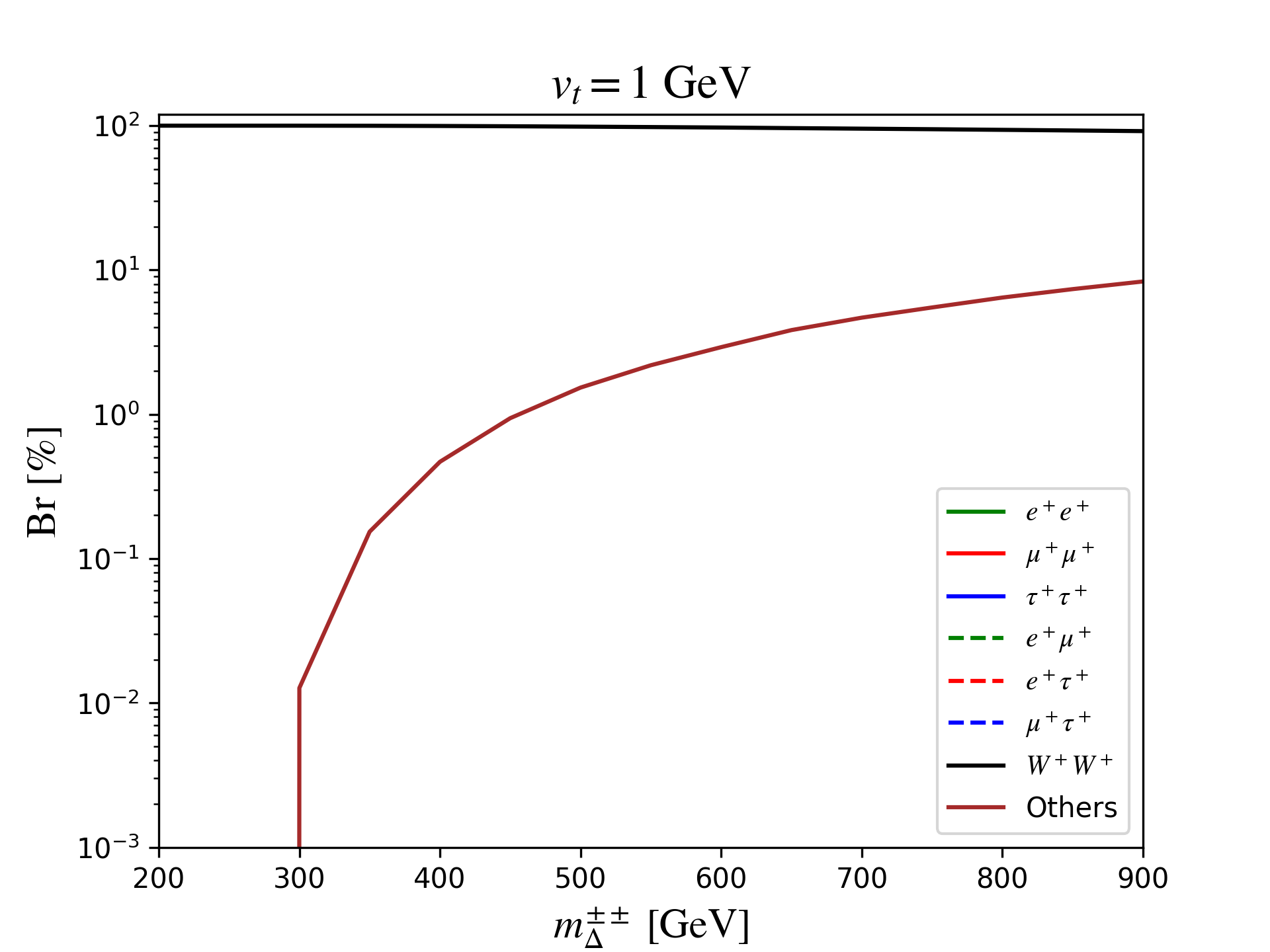}
    \caption{Branching ratios of the doubly charged scalar $\Delta^{\pm\pm}$ as functions of $m_{\Delta^{\pm\pm}}$ for two representative triplet vevs, $v_t=10^{-5}$~GeV (left) and $v_t=1$~GeV (right). The category ``Others'' denotes the sum of all remaining decay modes, including subleading three-body final states when kinematically allowed.}
    \label{fig:BR_mdel}
\end{figure*}

For such a degenerate spectrum, cascade transitions among the triplet states, such as $\Delta^{\pm\pm}\to\Delta^\pm W^{\pm(*)}$, are absent or phase-space suppressed. The phenomenology of the $\Delta^{\pm\pm}$ state is then controlled mainly by its direct decays into same-sign charged leptons or same-sign $W$ bosons. For on-shell $W$ bosons, the partial widths into vector bosons and charged leptons are respectively given by~\cite{Chun:2003ej, FileviezPerez:2008jbu, Melfo:2011nx}
\begin{equation}\begin{split}
  &\Gamma_{\Delta^{\pm\pm}\to W^\pm W^\pm} = \frac{g^4 v_t^2}{8 \pi m_{\Delta^{\pm\pm}}} \sqrt{1 \!-\! \left( \frac{2m_W}{m_{\Delta^{\pm\pm}}} \right)^2}\\[.2cm]
  &\hspace{2cm} \times \left[2 + \left( \frac{m_{\Delta^{\pm\pm}}^2}{2 m_W^2} -1 \right)^2 \right],\\[.2cm]
  &\Gamma_{\Delta^{\pm\pm}\to l_i^\pm l_j^\pm} = \frac{m_{\Delta^{\pm\pm}}}{8\pi(1+\delta_{ij})} \left| \frac{M_{\nu_{ij}}} {v_t} \right|^2,
\end{split}\label{eq:deltadecay}\end{equation}
where the first expression is applicable only for $m_{\Delta^{\pm\pm}}>2m_W$. Below this threshold, one or both $W$ bosons are off shell and the corresponding three-body or four-body decay widths must be used. For fixed neutrino masses, the leptonic widths scale as $v_t^{-2}$, while the diboson width scales as $v_t^2$. This opposite dependence produces a transition from a leptonic regime at small $v_t$ to a diboson regime at larger $v_t$. Equating the total leptonic width with the diboson width gives a characteristic transition value, typically $v_t^{\rm crit}\sim 10^{-4}-10^{-3}$ GeV for $m_{\Delta^{\pm\pm}}\sim{\cal O}(100)$~GeV and neutrino masses of order $0.05$~eV, with the precise value depending on the scalar mass and neutrino parameters. In the leptonic regime, the relative flavor branching fractions are controlled by the entries of the neutrino mass matrix and therefore depend on the neutrino mass ordering, the lightest neutrino mass and the values of the neutrino Majorana phases~\cite{Garayoa:2007fw, Kadastik:2007yd}. However, for large triplet vevs $v_t\sim{\cal O}(1)$~GeV, the diboson mode $\Delta^{\pm\pm}\to W^\pm W^\pm$ dominates.

Figure~\ref{fig:BR_mdel} illustrates this behavior by showing the branching ratios of the $\Delta^{\pm\pm}$ scalar as functions of $m_{\Delta^{\pm\pm}}$ for two representative triplet vevs $v_t=10^{-5}$ GeV and $v_t=1$ GeV. These two benchmark choices therefore lie on opposite sides of the transition between the leptonic and diboson-dominated regimes. In addition, the neutrino parameters entering $M_\nu$ are fixed to their PDG central values~\cite{ParticleDataGroup:2024cfk}, assuming normal ordering, $m_{\nu_1}=0$ and vanishing Majorana phases. The solid colored curves correspond to same-flavor dilepton modes, the dashed curves to mixed-flavor dilepton modes, the black curve to the diboson channel and the brown curve labelled `Others' to the sum of all remaining decay modes including the three-body channels in the $t \bar b W$, $WWZ$ and $WWh$ final states when kinematically allowed. For the chosen neutrino benchmark, the same-flavor dilepton channels give the largest leptonic contributions in the small-$v_t$ regime, while the mixed-flavor modes are comparatively suppressed. This hierarchy is, however, not generic and can change for different choices of the lightest neutrino mass or neutrino mass ordering. For $v_t=1$ GeV, the diboson channel dominates over the full mass range shown. In addition, for the benchmark configuration considered in this work, the total width is sufficiently large for the $\Delta^{\pm\pm}$ state to decay promptly. Long-lived or displaced signatures arise only in more restricted regions of the parameter space where both the leptonic and diboson partial widths are suppressed~\cite{Dev:2018sel, Antusch:2018svb}.

\subsection{Experimental limits on doubly charged scalars masses}
\label{subsec:experiment}

Before discussing LHC constraints, we recall that LEP2 searches for pair-produced doubly charged scalars already exclude masses close to the kinematic limit of the collider, with lower limits around $m_{\Delta^{\pm\pm}}\simeq 100$ GeV at 95\% confidence level in several benchmark interpretations~\cite{OPAL:2001luy, DELPHI:2002bkf, L3:2003zst}. Direct LHC bounds on the mass of the $\Delta^{\pm\pm}$ state depend strongly on the triplet vev, the scalar mass spectrum, the production mode and the dominant decay channel. In the small-$v_t$ regime where $\Delta^{\pm\pm}\to\ell^\pm\ell^\pm$ dominates, same-sign dilepton and multilepton searches provide the strongest constraints. The ATLAS Run-2 search with $139~{\rm fb}^{-1}$ at $\sqrt{s}=13$ TeV hence excludes doubly charged Higgs boson masses up to $m_{\Delta^{\pm\pm}}=1080$ GeV under the benchmark assumption of equal branching fractions into all charged-lepton flavor combinations~\cite{ATLAS:2022pbd}. Earlier CMS and ATLAS searches at $\sqrt{s}=13$ TeV also set stringent limits from the study of leptonic final states, though based on smaller integrated luminosities and yielding correspondingly weaker constraints~\cite{CMS:2017pet, ATLAS:2017iqw}. These leptonic limits, however, do not directly apply to the large-$v_t$ regime in which $\Delta^{\pm\pm}\to W^\pm W^\pm$ dominates.

For the diboson-dominated regime, an ATLAS search for pair-produced doubly charged scalars decaying as $\Delta^{\pm\pm}\to W^\pm W^\pm$, based on $36.1~{\rm fb}^{-1}$ at $\sqrt{s}=13$~TeV, excluded the benchmark model for $200~{\rm GeV}<m_{\Delta^{\pm\pm}}<220~{\rm GeV}$~\cite{ATLAS:2018ceg}. In addition, the stronger full Run-2 ATLAS search with $139~{\rm fb}^{-1}$ targets doubly and singly charged Higgs bosons decaying into electroweak gauge bosons in multilepton final states, including both pair production, $pp\to\Delta^{++}\Delta^{--}$ with $\Delta^{\pm\pm}\to W^\pm W^\pm$, and associated production, $pp\to\Delta^{\pm\pm}\Delta^\mp$ with $\Delta^{\pm\pm}\to W^\pm W^\pm$ and $\Delta^\mp\to W^\mp Z$. In the Type-II seesaw interpretation, the pair-production-only scenario excludes $\Delta^{\pm\pm}$ masses up to about $350~{\rm GeV}$, while the associated-production scenario yields a limit of about $230~{\rm GeV}$~\cite{ATLAS:2021jol}.

Finally, the vector boson fusion (VBF) production of doubly charged scalars is also possibly relevant because the coupling to electroweak gauge bosons grows with the triplet vev. However, in the minimal Type-II seesaw model, the triplet vev is constrained by electroweak precision data to be at most of order a few GeV. Therefore, VBF production does not provide the leading direct constraints relative to Drell-Yan pair and associated production searches~\cite{Agrawal:2018pci}. Moreover, recent VBF searches for singly and doubly charged scalar bosons decaying into vector bosons are mainly interpreted in the Georgi-Machacek framework, and cannot thus be directly translated into the minimal Type-II seesaw scenario considered here without a dedicated reinterpretation study~\cite{ATLAS:2024txt}. 

In this work, we focus on a nearly degenerate diboson-dominated Type-II seesaw scenario, thus with a large triplet vacuum expectation value $v_t\simeq {\cal O}(1)$~GeV. The diboson bounds discussed above are the most relevant direct LHC bounds, such that we consider $m_{\Delta^{\pm\pm}}=200-900$~GeV. This range covers not only the currently constrained low-mass region, but also masses beyond the present diboson-search exclusions satisfying $m_{\Delta^{\pm\pm}}\gtrsim350$~GeV. In such a parameter space region, direct LHC sensitivity remains limited so that $\mu$TRISTAN has the potential to provide a complementary probe through dedicated reconstruction and selection strategies.

\section{Doubly Charged Scalar Sensitivity at \texorpdfstring{$\mu$TRISTAN}{muTRISTAN}}
\label{sec:analysis}

\subsection{Signal topology and event simulation}\label{sec:topology}
We examine the production of a pair of doubly charged scalars at a muon collider operated in the $\mu^+\mu^+$ mode, targeting the large-$v_t$ diboson-dominated regime for which the leptonic searches lose sensitivity and the most relevant existing LHC constraints come from diboson searches. We take the center-of-mass energy $\sqrt{s}=2$~TeV as a representative high-energy $\mu^+\mu^+$ configuration, motivated by the original $\mu$TRISTAN proposal in which same-sign antimuon collisions up to this centre-of-mass energy are envisaged~\cite{Hamada:2022mua}. 

The multi-boson signal process considered is
\begin{equation}\begin{split}
  \mu^+\mu^+ \to&\ \mu^+\mu^+ \Delta^{++}\Delta^{--}  \\
    \to &\ \mu^+\mu^+ W^+W^+W^-W^- .
\end{split}\end{equation}
For large triplet vevs, this final state is produced mainly through gauge-mediated topologies in which the incoming muons scatter by radiating neutral electroweak bosons to produce the $\Delta^{++}\Delta^{--}$ pair. Our study, however, retains the full tree-level matrix element for the $\mu^+\mu^+\to\mu^+\mu^+\Delta^{++}\Delta^{--}$ process, including all contributing diagrams and their interferences. We design an analysis targeting the fully hadronic decay of the four $W$ bosons,
\begin{equation}
  \mu^+\mu^+   \to \mu^+\mu^+ \Delta^{++}\Delta^{--}   \to \mu^+\mu^+ + 8j ,
\end{equation}
which benefits from the large hadronic branching fraction of the $W$ boson, ${\rm Br}(W\to jj)\simeq 67\%$ corresponding to a fully hadronic branching fraction of about $[{\rm Br}(W\to jj)]^4\simeq 20\%$ for the four-$W$ bosons system, while presenting a non-trivial combinatorial reconstruction challenge. 

\begin{figure}
   \centering
   \includegraphics[width=0.95\columnwidth]{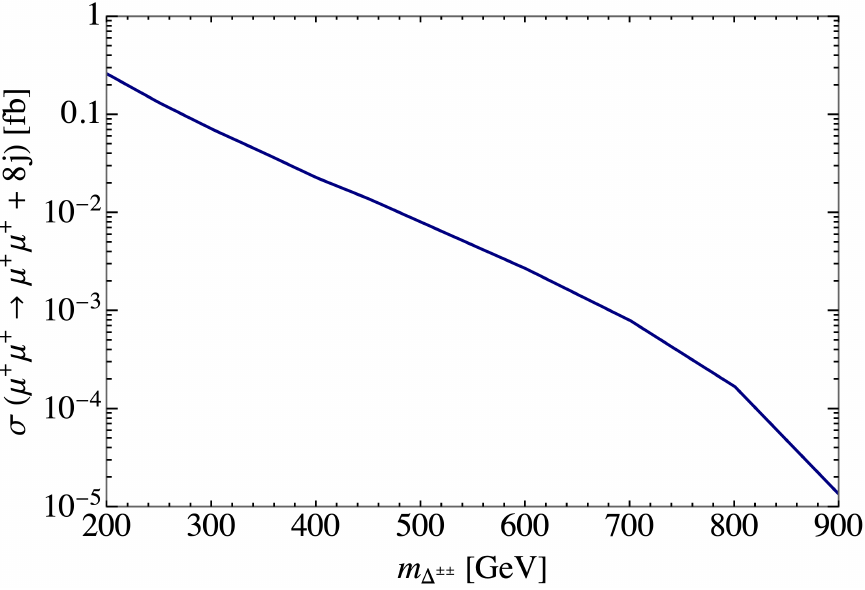}
   \caption{Signal cross section for the fully hadronic process $\mu^+\mu^+\to\mu^+\mu^+ \Delta^{++}\Delta^{--} \to \mu^+\mu^+ + 8j$ as a function of the doubly charged scalar mass $m_{\Delta^{\pm\pm}}$ at $\sqrt{s}=2$ TeV. The rates are computed from the full tree-level matrix element, include branching ratios for the $\Delta^{\pm\pm}\to W^\pm W^\pm$ and $W\to jj$ decays, and assume a benchmark scenario with $v_t=1$ GeV and a nearly degenerate triplet spectrum.}
    \label{fig:sigma_mdel}
\end{figure}

We assume a nearly degenerate triplet spectrum, $m_{\Delta^{\pm\pm}}=m_{\Delta^\pm}=m_{\Delta^0}=m_\chi$, and fix the triplet vev to $v_t=1$ GeV. The neutrino parameters entering the triplet Yukawa matrix are set to their PDG central values, assuming normal ordering, a vanishing lightest neutrino mass and zero Majorana phases~\cite{ParticleDataGroup:2024cfk}. In this large-$v_t$ regime, the leptonic branching fractions of the $\Delta^{\pm\pm}$ states are negligible, so the signal rates are practically insensitive to the properties of the neutrino sector. Finally, we consider representative doubly charged scalar masses in the range
\begin{equation}
  m_{\Delta^{\pm\pm}}\in [200, 900]~{\rm GeV} .
\end{equation}

To calculate the associated signal cross sections, we use the \textsc{TypeIISeesaw} UFO libraries~\cite{Fuks:2019clu} generated within the \textsc{FeynRules}/UFO framework~\cite{Christensen:2009jx, Alloul:2013bka, Degrande:2011ua, Darme:2023jdn}, which contain the Type-II seesaw scalar-triplet interactions relevant for the production and decay of the $\Delta^{\pm\pm}$ states. The corresponding leading-order parton-level rates, including the branching ratios for the $\Delta^{\pm\pm}\to W^\pm W^\pm$ and $W\to jj$ decays, are then computed with \textsc{MadGraph5\_aMC@NLO}~\cite{Alwall:2014hca} for a collider energy of \mbox{$\sqrt{s}=2$}~TeV. 

The results are shown in Figure~\ref{fig:sigma_mdel}. The fully hadronic signal rate decreases rapidly with increasing $m_{\Delta^{\pm\pm}}$ values, falling by several orders of magnitude over the mass range considered. This behavior reflects the shrinking phase space for producing the heavy scalar pair at fixed collider energy, together with the suppression of configurations in which the incoming muons radiate neutral electroweak bosons energetic enough to produce a high-invariant-mass $\Delta^{++}\Delta^{--}$ system, as expected from the effective-vector-boson picture of high-energy lepton collisions~\cite{Ruiz:2021tdt}. In the large-$v_t$ benchmark adopted here, the decay $\Delta^{\pm\pm}\to W^\pm W^\pm$ dominates, with a branching ratio exceeding 90\% across most of the mass range. Consequently, the signal rate is essentially controlled by the electroweak production cross section and by the hadronic branching fractions of the four $W$ bosons, rather than by the triplet Yukawa couplings or the details of the neutrino sector. For an integrated luminosity of $1~{\rm ab}^{-1}$, the cross sections shown in Figure~\ref{fig:sigma_mdel} translate into sizable raw event yields at low masses. By contrast, the moderate-mass and high-mass regions become statistically challenging and therefore motivate a dedicated analysis strategy. 

We now turn to the event simulation tool chain used in our collider analysis. Signal and background hard-scattering events are generated at leading order with \textsc{MadGraph5\_aMC@NLO}, using the same UFO implementation as for the signal-rate calculation. The generated events are passed to \textsc{Pythia 8.3}~\cite{Bierlich:2022pfr} for parton showering and hadronization, and detector effects are simulated with \textsc{Delphes 3.5.0}~\cite{deFavereau:2013fsa} using a detector card adapted to the $\mu$TRISTAN multi-TeV muon-collider environment. In this framework, jets are reconstructed with the $k_T$ algorithm~\cite{Catani:1993hr, Ellis:1993tq} with radius parameter $R=0.4$, as implemented in \textsc{FastJet}~\cite{Cacciari:2011ma}. The use of a sequential-recombination algorithm follows the lepton-collider-oriented reconstruction implemented in typical relevant \textsc{Delphes} cards, since in the absence of LHC-like pile-up and underlying-event activity, such algorithms provide a suitable basis for resolving multijet final states. A complete treatment of these effects, including in particular the beam-induced backgrounds relevant for a lepton collider, would however require a dedicated detector-level study~\cite{MuonCollider:2022ded}, which lies beyond the scope of this work. 

Reconstructed charged leptons and jets are required to satisfy standard transverse momentum and pseudo-rapidity requirements,
\begin{equation}\label{eq:preselectiona}\begin{split}
  p_T^\ell>10~{\rm GeV},\quad &|\eta^\ell|<2.5,\\
  p_T^j>25~{\rm GeV},\quad &|\eta^j|<2.5 .
\end{split}\end{equation}
As a loose baseline preselection, we require selected events to feature at least two reconstructed antimuons and at least two reconstructed jets,
\begin{equation}\label{eq:preselectionb}
  N_{\mu^+}\geq 2
  \qquad\text{and}\qquad
  N_j\geq 8 .
\end{equation}
The antimuon requirement reflects the same-sign scattering topology of the signal process, while the jet requirement ensures that the event contains enough hadronic activity to initiate a dijet-based W-boson reconstruction. More restrictive selections exploiting the presence of four hadronically decaying $W$ bosons in the signal final state will be introduced in Section~\ref{subsec:Cutbased_analysis}.

The dominant backgrounds arise from SM same-sign antimuon scattering in association with either QCD radiation or charged weak boson production. We generate inclusive background samples for the processes
\begin{equation}\begin{split}
  &\mu^+\mu^+ \to \mu^+\mu^+ + nj  \quad\text{with}\quad n=2,4,\\
  &\mu^+\mu^+ \to \mu^+\mu^+ W^+ W^-,
\end{split}\end{equation}
at the matrix-element level. These processes can mimic the signal when jets from QCD radiation are accidentally paired into dijet candidates with invariant masses close to the $W$-boson mass. Additional backgrounds with larger matrix-element parton multiplicities are expected to be suppressed by the corresponding phase space factors and extra powers of the strong coupling constant. In particular, we explicitly checked that the inclusive pure-QCD-radiation background $\mu^+\mu^+\to\mu^+\mu^+ +6j$ cross section before any analysis cuts is of order of $1~{\rm ab}$, making it negligible for the present sensitivity estimates. Similarly, the process $\mu^+\mu^+\to\mu^+\mu^+W^+W^-+2j$ has a preselection cross section of about $0.5~{\rm fb}$ before applying the $W\to jj$ branching fractions and any preselection cuts, and is therefore subleading with respect to inclusive $\mu^+\mu^+W^+W^-$ production.

For the processes considered, additional softer and collinear radiation is modeled through the parton shower in \textsc{Pythia 8.3}. Since the different jet-multiplicity samples are generated inclusively and are not matched or merged, some overlap can arise between hard radiation described by the parton shower in the lower-multiplicity samples and matrix-element radiation in the higher-multiplicity samples. For simplicity, we retain these contributions independently in our baseline background estimate, which thus gives a conservative treatment for the present analysis. Given the two order of magnitude hierarchy among the relevant multipartonic cross sections, this approximation is not expected to affect the qualitative sensitivity conclusions, while a fully exclusive simulation is left for future refinement.

At generator level, the inclusive cross sections are $246.13~{\rm fb}$ for the $\mu^+\mu^+\to\mu^+\mu^++2j$ process, $1.63~{\rm fb}$ for the $\mu^+\mu^+\to\mu^+\mu^++4j$ process and $68.43~{\rm fb}$ for the $\mu^+\mu^+\to\mu^+\mu^+W^+W^-$ process, including the hadronic branching fractions of the two $W$ bosons,
\begin{equation}\begin{split}
    \sigma(\mu^+\mu^+\to\mu^+\mu^++2j) =&\ 246.13~\mathrm{fb},\\
    \sigma(\mu^+\mu^+\to\mu^+\mu^++4j) = &\ 1.63~\mathrm{fb},\\
    \sigma(\mu^+\mu^+\to\mu^+\mu^+W^+W^-) = &\ 68.43~\mathrm{fb}.
\end{split}\end{equation}
However, after imposing the preselection of Eqs.~\eqref{eq:preselectiona} and \eqref{eq:preselectionb}, the corresponding fiducial rates are reduced to about $1.23~{\rm ab}$, $3.23~{\rm ab}$ and $1.7~{\rm ab}$, respectively. Thus, after the high jet-multiplicity requirement, the $\mu^+\mu^++4j$ and $\mu^+\mu^+W^+W^-$ contributions to the background provide the most signal-like topologies with multiple possible dijet pairings. The signal, by contrast, contains four hadronically decaying $W$ bosons and therefore a characteristic high-multiplicity resonant structure which we can further exploit to improve the analysis.

\begin{figure*}
    \centering
    \includegraphics[width=0.49\textwidth]{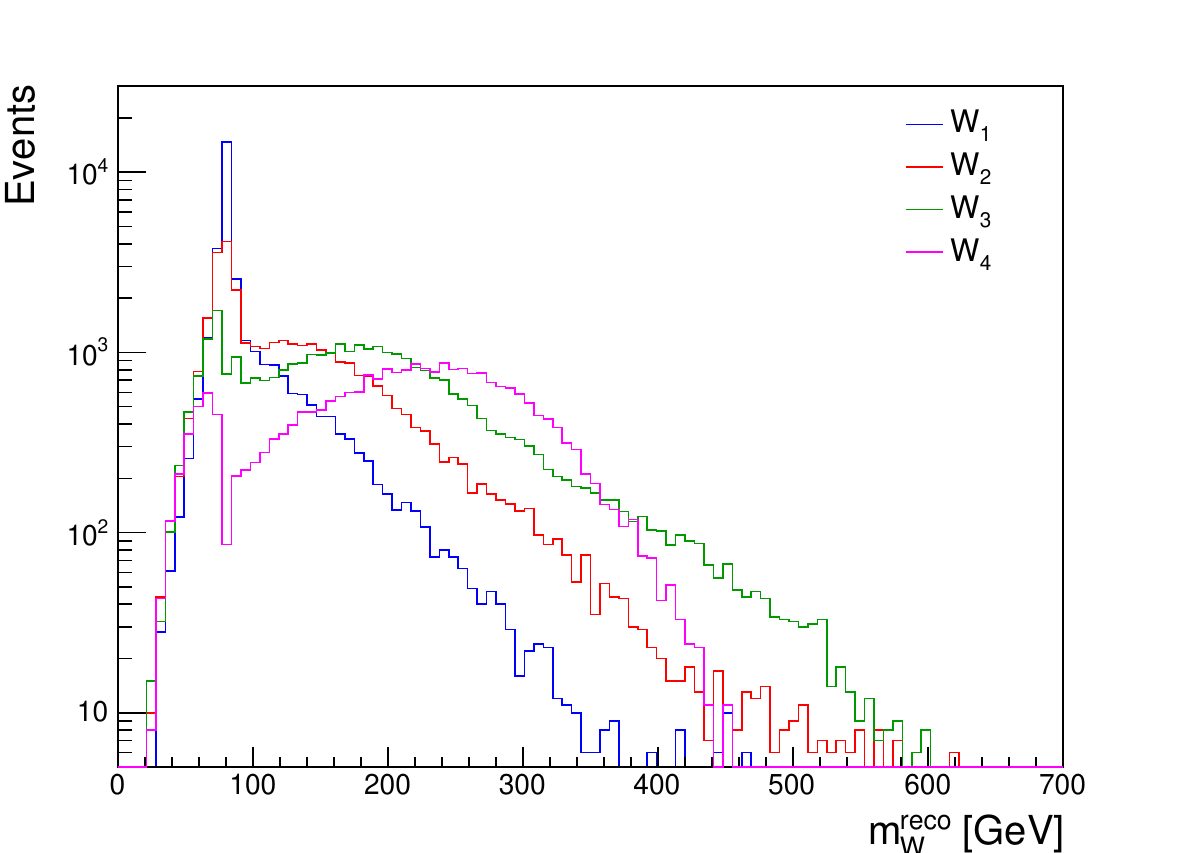}\hfill
    \includegraphics[width=0.49\textwidth]{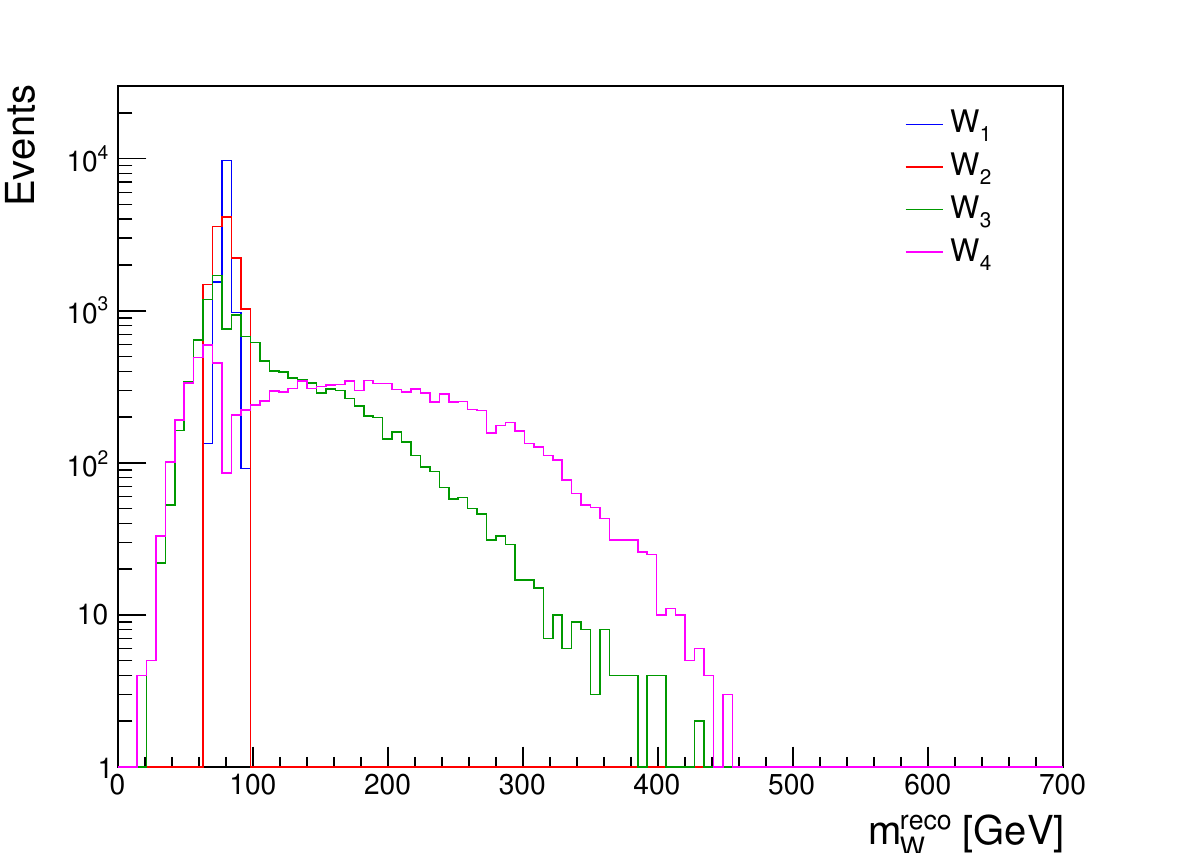}
    \caption{Distributions of the reconstructed dijet invariant masses for the four $W$ candidates in a representative signal benchmark for which $m_{\Delta^{\pm\pm}}=500~{\rm GeV}$. The left panel shows the distributions after the baseline and high-multiplicity requirements, while the right panel shows them after the $2tW$ selection. The $W$-boson candidates are ordered according to their individual $\chi^2_W$ values: $W_1$ denotes the best reconstructed candidate (blue), followed in sequence by $W_2$ (red), $W_3$ (green) and $W_4$ (magenta).}
    \label{fig:histo_2tW}
\end{figure*}

\subsection{Hadronic \texorpdfstring{$W$}{W} reconstruction and cut-based analysis}
\label{subsec:Cutbased_analysis}

After the baseline object selection and the high-multiplicity requirement discussed in Section~\ref{sec:topology}, we reconstruct the hadronically decaying $W$ bosons from the collection of reconstructed jets. For each event, we use the eight leading jets and group them into all possible configurations of four non-overlapping dijet pairs, corresponding to 105 distinct pairing assignments. For each configuration, we define
\begin{equation}
  \chi^2_{\rm tot} = \sum_{i=1}^{4} \chi^2_{W_i} = \sum_{i=1}^{4} \left( \frac{m_{W_i}^{\rm reco}-m_W}{\sigma_W} \right)^2 ,
\end{equation}
where $m_{W_i}^{\rm reco}$ is the invariant mass of the $i$-th dijet pair and $m_W=80.4~{\rm GeV}$. We take $\sigma_W=12~{\rm GeV}$ as an effective dijet mass resolution, motivated by the width of the reconstructed $W$ peak observed in our signal simulations. This fixed value is then used for all signal masses as a simplified, uniform reconstruction criterion, sufficient for defining the relative quality of the dijet pairings. The configuration with the smallest value of $\chi^2_{\rm tot}$ is then selected as the reconstructed event topology, thereby resolving the combinatorial ambiguity through global minimization. 

Within this globally selected configuration, the four resulting dijet candidates are ordered according to their individual $\chi^2_W$ values,
\begin{equation}
  \chi^2_{W_1}<\chi^2_{W_2}<\chi^2_{W_3}<\chi^2_{W_4} ,
\end{equation}
and four reconstructed $W$ candidates are retained, corresponding to the fully hadronic four-$W$ topology expected for the signal. The left panel of Figure~\ref{fig:histo_2tW} shows the reconstructed dijet masses for a representative signal benchmark with $m_{\Delta^{\pm\pm}}=500~{\rm GeV}$. Before imposing any requirement on the reconstructed $W$ candidates, the distributions already show a clear hierarchy among the four ordered candidates. The first two candidates $W_1$ and $W_2$ display the most pronounced accumulation around the physical $W$-boson mass. This is a direct consequence of the global $\chi^2_{\rm tot}$ minimization which favors pairing configurations in which several dijet masses lie close to $m_W$. The distributions of the third and fourth candidates are instead broader, reflecting the increasing impact of combinatorial ambiguities once the most $W$-like jet pairs have been selected.

Based on this hierarchy, we define three signal regions with increasing purity. The first region, denoted $2tW$, requires the two best reconstructed candidates to be compatible with a hadronic $W$ decay according to
\begin{equation}
  \chi^2_{W_1},\chi^2_{W_2} < \chi^2_{\rm tight}
  \qquad\text{with}\qquad
  \chi^2_{\rm tight}=2 .
\end{equation}
The second region, denoted $2t1lW$, keeps the same two tight-$W$ requirements and additionally demands a third, more loosely reconstructed candidate,
\begin{equation}\begin{split}
  &\chi^2_{W_1},\chi^2_{W_2}<\chi^2_{\rm tight}, \\
  &\chi^2_{W_3}<\chi^2_{\rm loose}
  \qquad\text{with}\qquad
  \chi^2_{\rm loose}=5 .
\end{split}\end{equation}
Finally, the most restrictive region, denoted $3tW$, requires three tight reconstructed $W$ candidates,
\begin{equation}
  \chi^2_{W_1},\chi^2_{W_2},\chi^2_{W_3} < \chi^2_{\rm tight} .
\end{equation}
These regions are not exclusive; rather, they are nested working points with increasing reconstruction purity, $3tW \subset 2t1lW \subset 2tW$. They should therefore be interpreted as alternative selections rather than orthogonal categories to be combined. Moreover, the numerical $\chi^2$ thresholds are chosen as representative working points balancing signal efficiency and background rejection. With $\sigma_W=12~{\rm GeV}$, the tight threshold $\chi^2_{\rm tight}=2$ corresponds to $|m_W^{\rm reco}-m_W|\lesssim17~{\rm GeV}$, while the loose threshold $\chi^2_{\rm loose}=5$ corresponds to \mbox{$|m_W^{\rm reco}-m_W|\lesssim 27~{\rm GeV}$}.

\begin{table*}\setlength{\tabcolsep}{12pt}
  \centering
  \renewcommand{\arraystretch}{1.45}
  \begin{tabular}{lccccc}
    Background process & $\sigma_{\rm gen}$ [fb] & $\varepsilon_{\rm base}$ [\%] & $\varepsilon_{2tW}$ [\%] & $\varepsilon_{2t1lW}$ [\%] & $\varepsilon_{3tW}$ [\%] \\
    \hline
    $\mu^+\mu^+\to\mu^+\mu^+ +2j$ & $246.13$ & $0.0005$ & $\sim 0$ & $\sim 0$ & $\sim 0$ \\
    $\mu^+\mu^+\to\mu^+\mu^+ +4j$ & $1.63$ & $0.198$ & $0.070$ & $0.044$ & $0.014$ \\
    $\mu^+\mu^+\to\mu^+\mu^+ W^+W^-$ & $68.43$ & $0.0055$ & $0.0011$ & $\sim0$ & $\sim0$\\
    \hline
    Total background rate [ab] & $-$ & $8.22$ & $1.89$ & $0.72$ & $0.23$ \\
  \end{tabular}
  \caption{Background efficiencies and corresponding rates for the baseline selection ($\varepsilon_{\rm base}$) and for the three reconstructed-$W$ regions ($\varepsilon_{2tW}$, $\varepsilon_{2t1lW}$ and $\varepsilon_{3tW}$). For the $W^+W^-$ background, the quoted generated cross section already includes the two hadronic $W$-decay branching fractions, with the efficiencies thus describing the subsequent analysis selection. The last row gives the total background rate after each selection.\label{tab:bkg_eff_cut}}\vspace{.3cm}

  \resizebox{\linewidth}{!}{\begin{tabular}{c | c c | c c | c c| c c}
    $m_{\Delta^{\pm\pm}}$ [GeV] & $\sigma_{\rm sig}$ [fb] & $\varepsilon_{\rm base}$ [\%] & $\varepsilon_{2tW}$ [\%] & ${\cal S}_{2tW}$ & $\varepsilon_{2t1lW}$ [\%] & ${\cal S}_{2t1lW}$ & $\varepsilon_{3tW}$ [\%] & ${\cal S}_{3tW}$ \\
    \hline
    $200$ & $0.262$ & $12.67$ & $8.99$ & $4.668$ & $7.05$ & $4.217$ & $5.25$ & $3.680$ \\
    $300$ & $0.072$ & $34.52$ & $21.30$ & $3.693$ & $14.99$ & $3.181$ & $10.92$ & $2.760$ \\
    $400$ & $0.023$ & $42.18$ & $22.64$ & $1.952$ & $14.31$ & $1.644$ & $10.72$ & $1.500$ \\
    $500$ & $8.1\times10^{-3}$ & $42.18$ & $19.74$ & $0.853$ & $11.12$ & $0.708$ & $8.53$ & $0.721$ \\
    $700$ & $8.02\times10^{-4}$ & $29.49$ & $11.47$ & $0.065$ & $4.77$ & $0.044$ & $3.50$ & $0.055$ \\
    $900$ & $1.34\times10^{-5}$ & $12.89$ & $8.01$ & $7.76\times10^{-4}$ & $3.85$ & $6.09\times10^{-4}$ & $2.77$ & $7.76\times10^{-4}$ \\
  \end{tabular}}
  \caption{Signal efficiencies and expected significances for representative doubly charged scalar masses and a luminosity ${\cal L}=1~{\rm ab}^{-1}$, for the baseline selection and for the three reconstructed-$W$ regions.}
  \label{tab:sig_off_cut}
\end{table*}

In the right panel of Figure~\ref{fig:histo_2tW}, we illustrate the impact of imposing the $2tW$ requirement for a benchmark configuration with $m_{\Delta^{\pm\pm}}=500~{\rm GeV}$. By construction, the first two $W$-boson candidates are sharply localised around $m_W$. In addition, the remaining candidates still retain information about the four-$W$ bosons nature of the signal. While the $W_3$ and $W_4$ distributions remain broader than the $W_1$ and $W_2$ ones, they indeed retain a visible accumulation near the $W$-boson mass, which motivates the use of additional reconstruction regions requiring a third loosely or tightly reconstructed $W$ candidate. The $2tW$ region is therefore designed to maximise the signal efficiency, while the $2t1lW$ and $3tW$ regions trade part of this efficiency for increased purity by probing the higher-multiplicity resonant structure characteristic of the signal.

For each selection region $i={\rm base}$, $2tW$, $2t1lW$ and $3tW$, where the baseline region includes only the object and jet-multiplicity requirements discussed in Section~\ref{sec:topology}, we compute the surviving number of events $N_i$ for a benchmark integrated luminosity ${\cal L}=1~{\rm ab}^{-1}$. The corresponding efficiencies are defined as
\begin{equation}
    \varepsilon_i = \frac{N_i}{N_0} ,
\end{equation}
where $N_0$ denotes the number of events prior to any selection cut. The impact of the baseline and of the reconstructed-$W$ selections on the background samples is shown in Table~\ref{tab:bkg_eff_cut}. As already mentioned, after the high jet-multiplicity requirement all backgrounds are reduced to the ab level. The pure $\mu^+\mu^++2j$ sample, although very large inclusively, is almost entirely rejected by the requirement of eight reconstructed jets. The $\mu^+\mu^++4j$ background is smaller at generator level but has a larger probability to populate the high-multiplicity region through additional shower radiation. The genuine $\mu^+\mu^+W^+W^-$ background is also strongly suppressed by the eight-jet requirement, but remains relevant because its hadronic weak-boson decays naturally generate dijet pairs with an invariant mass close to the $W$ mass. The additional $2tW$, $2t1lW$ and $3tW$ restrictions further reduce the background rate by requiring the presence of reconstructed $W$ candidates.

The expected signal and background yields in each reconstructed-$W$ region are, respectively, obtained as
\begin{equation}
  S_i = {\cal L}\,\sigma_{\rm sig}\,\varepsilon_i^{\rm sig}
  \qquad\text{and}\qquad
  B_i = {\cal L}\, \sum_b \sigma_b\,\varepsilon_i^b ,
\end{equation}
where $i=2tW$, $2t1lW$ and $3tW$, and the background sum runs over the background components. We then compute the significance
\begin{equation}\label{eq:significance}
  {\cal S}_i = \frac{S_i}{\sqrt{S_i+B_i}} ,
\end{equation}
which provides a standard measure of the expected sensitivity for a counting experiment. The results for both the efficiencies and the sensitivity are shown in Table~\ref{tab:sig_off_cut}. The signal efficiencies exhibit a non-trivial dependence on the doubly charged scalar mass. After the baseline high-multiplicity requirement, the efficiency rises from about $13\%$ at $m_{\Delta^{\pm\pm}}=200~{\rm GeV}$ to about $40\%$ in the intermediate-mass region around $400-500~{\rm GeV}$, before decreasing again at larger masses. This behaviour reflects that at low masses, some of the jets produced in the $W$ decays can fail the transverse-momentum requirement of Eq.~\eqref{eq:preselectiona}, whereas at high masses the four-$W$ system becomes more energetic and the resolved small-radius reconstruction gets increasingly affected by the more boosted event topology and by the combinatorial ambiguities.

\begin{figure*}
    \centering
    \includegraphics[width=0.47\textwidth]{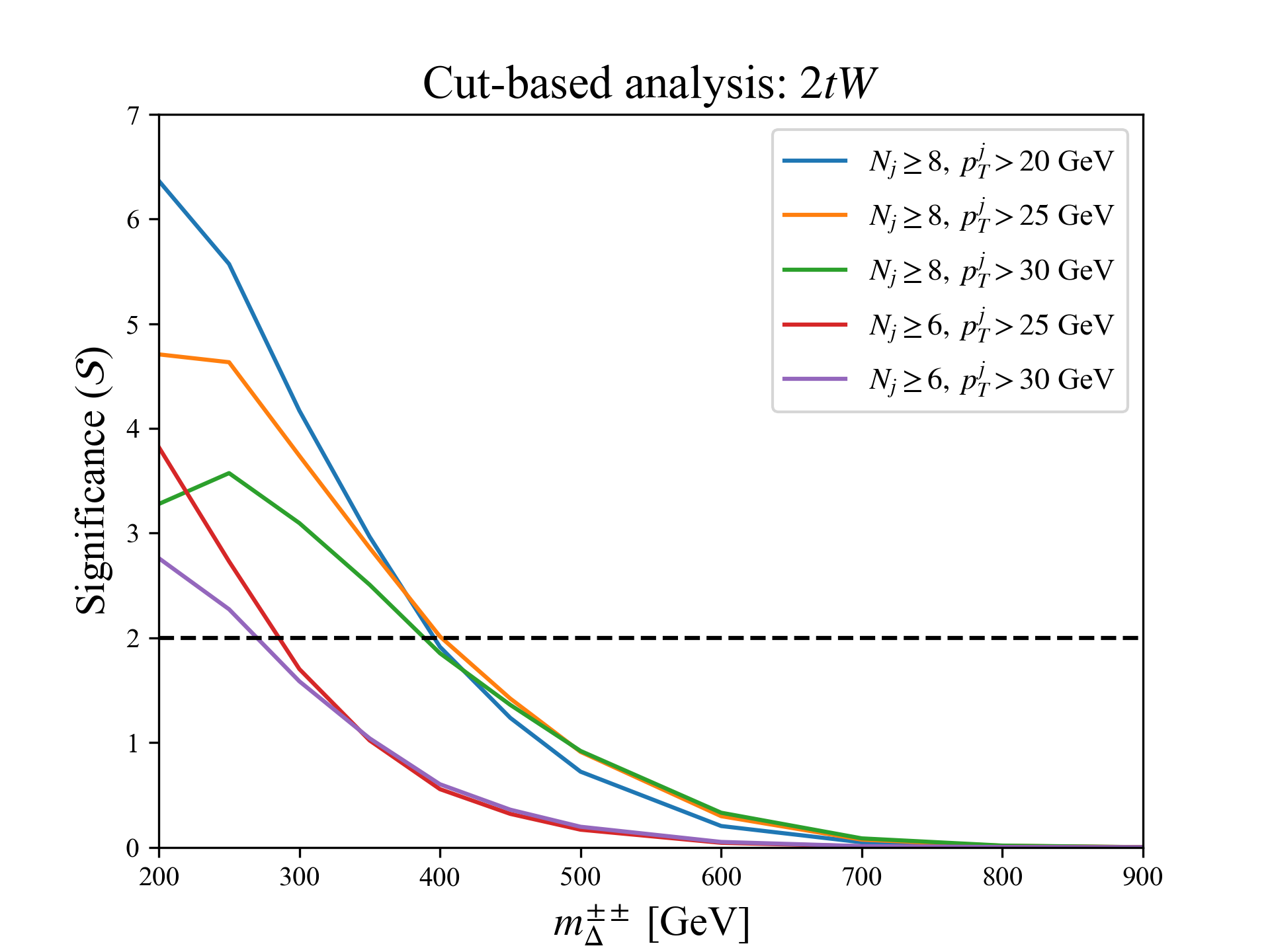}\hfill
    \includegraphics[width=0.47\textwidth]{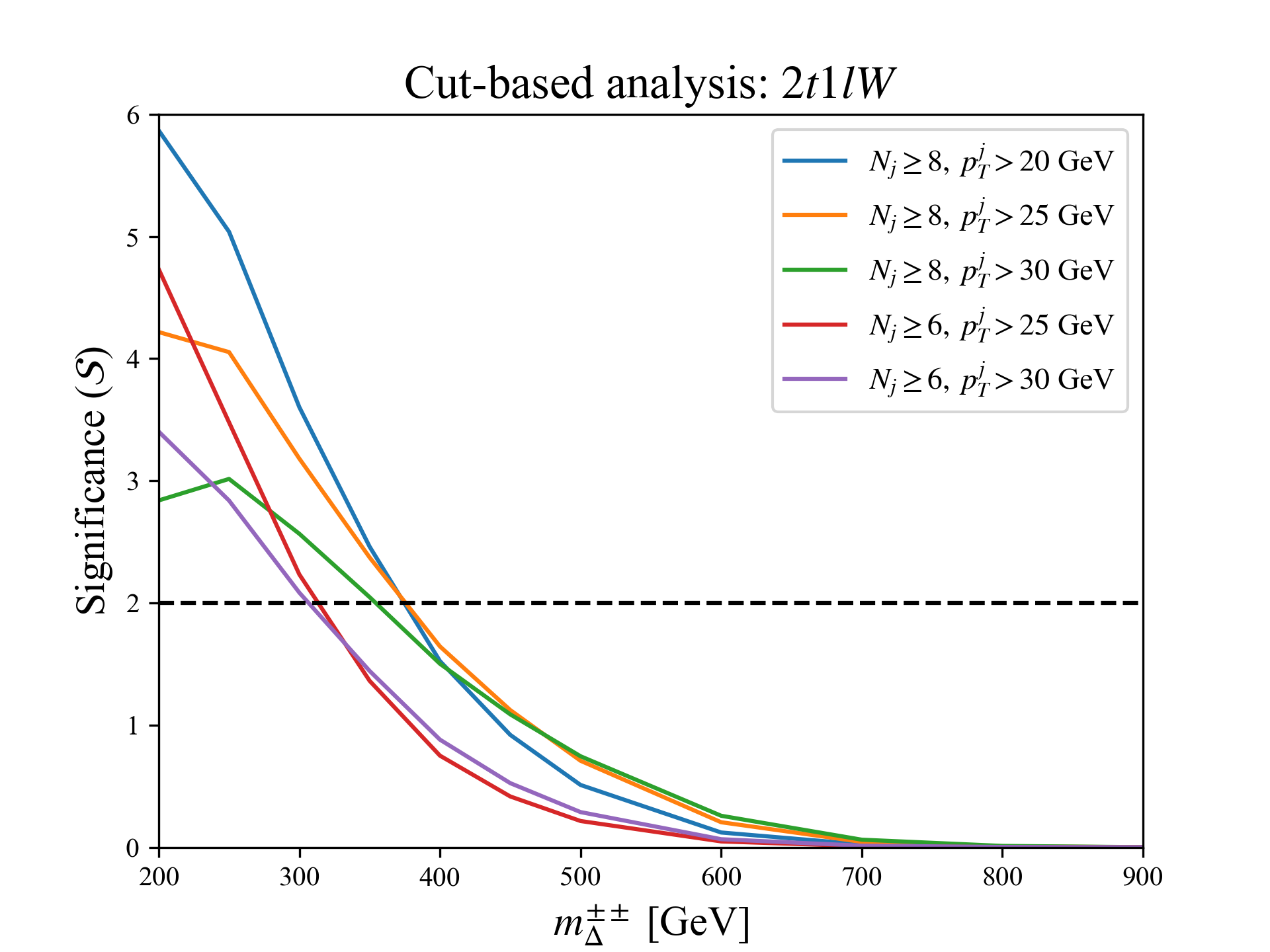}\\
    \includegraphics[width=0.47\textwidth]{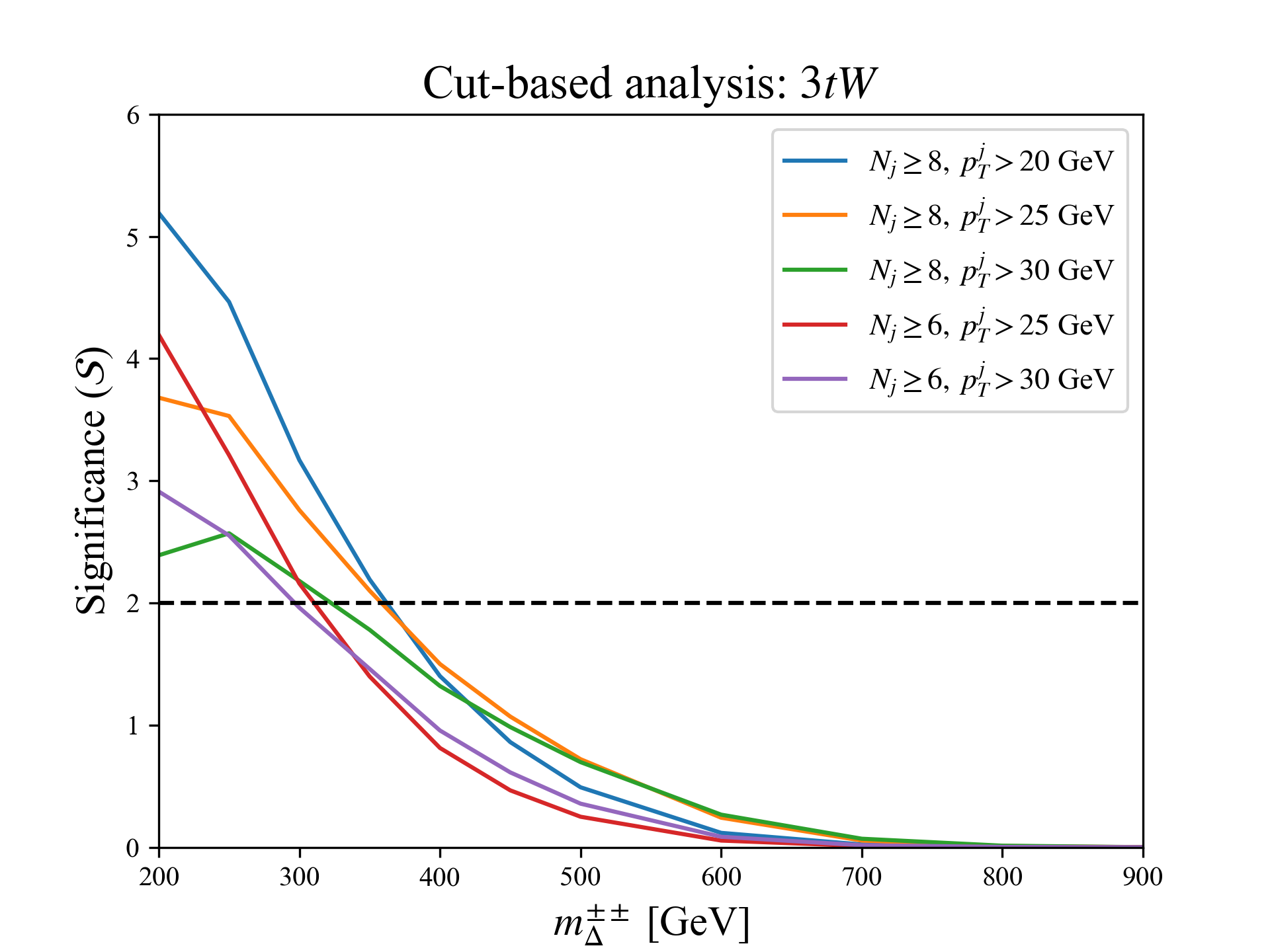}
    \caption{Expected cut-based significance as a function of $m_{\Delta^{\pm\pm}}$ for different choices of the jet-multiplicity and jet-$p_T$ requirements. The three panels correspond to the $2tW$ (top left), $2t1lW$ (top right) and $3tW$ (bottom) reconstruction regions. The dashed horizontal line indicates ${\cal S}=2$, and the significances are evaluated for ${\cal L}=1~{\rm ab}^{-1}$.}
    \label{fig:cutbased}
\end{figure*}

The $2tW$ region provides the best cut-based sensitivity over most of the mass range, as displayed in Table~\ref{tab:sig_off_cut}. This is because the baseline selection already reduces the backgrounds to the few-event level, so that requiring only two tightly reconstructed $W$ candidates retains a large fraction of the signal while providing sufficient background rejection. The more restrictive $2t1lW$ and $3tW$ regions reduce the backgrounds further, but the gain in purity does not compensate for the loss in signal efficiency in a simple cut-and-count analysis. They nevertheless provide useful higher-purity working points, as they probe the presence of more than two resonant hadronic $W$ candidates, a feature absent or strongly suppressed in the dominant backgrounds. Numerically, the cut-based strategy exceeds evidence-level sensitivity for light doubly charged scalars. For $m_{\Delta^{\pm\pm}}=200-300~{\rm GeV}$, the $2tW$ region yields expected significances of about $3.7-4.7\sigma$. The sensitivity decreases rapidly at larger masses, falling below $2\sigma$ around $m_{\Delta^{\pm\pm}}\simeq400~{\rm GeV}$, and well below $2\sigma$ beyond this range. This loss of sensitivity is, however, driven primarily by the steep decrease of the production cross section with increasing scalar mass, rather than by a failure of the reconstruction efficiencies alone.

We estimate the robustness of these conclusions against background normalization uncertainties by replacing the significance introduced in Eq.~\eqref{eq:significance} by
\begin{equation}
  {\cal S}_i(\delta_B) = \frac{S_i}{\sqrt{S_i+B_i+(\delta_B B_i)^2}} ,
\end{equation}
where $\delta_B$ denotes an effective relative uncertainty on the total background yield in the considered region. Since the selected backgrounds are very small, with $B_{2tW}\simeq1.9$ events and even smaller in the $2t1lW$ and $3tW$ regions, conventional normalization uncertainties of $\delta_B=5\%$ or $10\%$ have a negligible impact on the quoted significances. As a more conservative stress test for background modeling, we also rescale the total background yield by a factor $\kappa_B=2$ or 3 and compute
\begin{equation}
  {\cal S}_i(\kappa_B) = \frac{S_i}{\sqrt{S_i+\kappa_B B_i}} .
\end{equation}
This test is not meant to replace a dedicated assessment of detector and reconstruction systematics, but it approximately brackets the impact of large residual background-modeling uncertainties in the highly suppressed tails of the analysis. In particular, it is expected to cover the intrinsic normalization uncertainty of the leading-order background predictions, the absence of matching or merging in the multijet samples, as well as possible shower-modeling effects. In the most sensitive $2tW$ region, the significance of $m_{\Delta^{\pm\pm}}=200~{\rm GeV}$ changes from ${\cal S}=4.67$ to 4.51 and 4.36 when the background is multiplied by factors of two and three, respectively. For $m_{\Delta^{\pm\pm}}=300~{\rm GeV}$, the corresponding significance values are ${\cal S}=3.70$, 3.51 and 3.35, illustrating that the low-mass sensitivity is robust even under sizable background rescalings. At larger masses, the significance becomes more background-sensitive, but the dominant limitation remains the rapidly falling signal production rate.

We have also varied the jet-multiplicity and jet-$p_T$ requirements of Eqs.~\eqref{eq:preselectiona} and~\eqref{eq:preselectionb} to assess the robustness of the baseline choice. The corresponding impact on the cut-based significance is shown in Figure~\ref{fig:cutbased} for the three reconstructed-$W$ regions. Lowering the jet threshold to $p_T^j>20~{\rm GeV}$ increases the signal acceptance and leads to a larger significance in the low-mass region, but raises the total background rate after the baseline requirement from about $8.2~{\rm ab}$ to about $13.4~{\rm ab}$. Conversely, raising the threshold to $p_T^j>30~{\rm GeV}$ reduces the baseline background to about $2.6~{\rm ab}$, but at the price of a lower signal efficiency, especially for lighter scalar masses where some jets from the $W$ decays are softer. Relaxing the multiplicity requirement has a more dramatic effect. For instance, with $N_j\geq6$ and $p_T^j>30~{\rm GeV}$, the baseline background increases to about $0.68~{\rm fb}$, driven mainly by the genuine $\mu^+\mu^+W^+W^-$ contribution, while for $N_j\geq4$ (not shown in the figure) it reaches the ten-fb level. The high-multiplicity requirement $N_j\geq8$ is therefore essential to isolate the fully hadronic four-$W$ topology and to suppress backgrounds with only two genuine hadronic $W$ bosons. The previously discussed baseline setup in which we impose $N_j\geq8$ for a transverse-momentum threshold $p_T^j>25~{\rm GeV}$ provides a conservative compromise between the signal acceptance, the background rejection and the robustness of the resolved reconstruction.

Finally, it is useful to compare these results with the existing LHC bounds discussed in Section~\ref{sec:model}. Current LHC searches in the diboson regime constrain doubly charged scalars with masses up to the few-hundred-GeV range, with the precise value depending on the production mode, the doubly charged scalar decay pattern and the triplet spectrum. The cut-based $\mu$TRISTAN strategy developed here hence probes a comparable mass range, with evidence-level sensitivity for $m_{\Delta^{\pm\pm}}\lesssim 300-350~{\rm GeV}$ at ${\cal L}=1~{\rm ab}^{-1}$. This comparison shows that same-sign muon collisions provide a complementary probe of the large-$v_t$ diboson-dominated regime of the Type-II seesaw framework, independently of the LHC proton-proton environment. Our cut-based analysis therefore establishes the viability of the fully hadronic four-$W$ strategy, but it also highlights the limitations of relying only on reconstructed-$W$ counting variables. Extending the sensitivity beyond the range already probed by current diboson searches requires exploiting more of the event kinematics and correlations than is captured by the three simple reconstruction regions. This motivates the multivariate analysis developed in the next subsection.

\subsection{Multivariate analysis of the four-\texorpdfstring{$W$}{W} topology}\label{subsec:BDT_analysis}

The cut-based analysis presented in Section~\ref{subsec:Cutbased_analysis} demonstrates that the fully hadronic four-$W$ topology can be efficiently exploited, but it also relies on a small number of properties of the reconstructed $W$ bosons and therefore uses only part of the available event information. To recover additional sensitivity, especially in the region where the signal production rate becomes small, we complement this study with a multivariate analysis based on boosted decision trees (BDTs). In contrast to sequential cut-based selections, BDTs can exploit nonlinear correlations among observables and thus combine information, in this case, from the reconstructed $W$ candidates, the global event activity, and the angular and kinematic properties of the final state~\cite{Friedman:2001, Roe:2004na, Hastie:2009itz, Cornell:2021gut, Choudhury:2024crp, Cornell:2024dki}. Our implementation uses the \textsc{XGBoost} package~\cite{Chen:2016btl}, which provides an efficient and regularized realization of gradient-boosted decision trees. For both signal and background samples, we apply the same baseline object selection and preselection requirements as in Section~\ref{subsec:Cutbased_analysis}, thus ensuring a direct comparison with the cut-based strategy. A separate classifier is trained for each benchmark value of $m_{\Delta^{\pm\pm}}$, allowing the BDT response to adapt to the mass-dependent kinematics of the signal while keeping the background definition fixed across the scan.

To train the BDT classifiers, we construct a set of event-level observables designed to capture both the quality of the $W$ candidates and the global structure of the event. We consider two jet-multiplicity setups. The first setup follows the baseline strategy of Section~\ref{subsec:Cutbased_analysis}, summarized by Eqs.~\eqref{eq:preselectiona} and~\eqref{eq:preselectionb}, thus requiring at least eight reconstructed jets and targeting the fully resolved four-$W$ topology. The second setup relaxes this requirement to $N_j\geq6$ in order to assess whether the multivariate analysis can recover signal events in which part of the hadronic activity is softer, partially merged or otherwise not reconstructed as eight selected jets. As shown in Section~\ref{subsec:Cutbased_analysis}, this substantially increases the background rate. This setup therefore provides a useful test of whether the BDT can recover some of the discrimination lost by loosening the jet-multiplicity requirement, using additional kinematic and reconstruction-quality information. For each multiplicity setup, we also vary the jet-$p_T$ threshold around the baseline value $p_T^j>25~{\rm GeV}$. 

The input variables include the masses, the transverse momenta and the pseudorapidities of the reconstructed $W$ bosons, as well as the corresponding reconstruction-quality variables and the angular separations between the two constituent jets,
\begin{equation}
    \Big\{ m_{W_i}^{\rm reco}, \quad p_T(W_i), \quad \eta(W_i) ,\quad \chi^2_{W_i},  \quad \Delta R_{jj}^{(i)} \Big\} .
\end{equation}
In addition, we also consider the cosine of the decay angle of the leading jet in the rest frame of its parent $W$ candidate, 
\begin{equation}
    \Big\{  \cos\theta^*_{W_i} \Big\} ,
\end{equation}
which allows to probe the internal decay geometry of the reconstructed $W_i$ candidate and can help distinguish genuine hadronic $W$ decays from accidental dijet combinations. For the $N_j\geq6$ setup, the variables associated with the three best reconstructed candidates are used ($i=1, 2, 3$), while for the $N_j\geq8$ setup we retain the four $W$ candidates ($i=1, 2, 3, 4$).

A second class of observables describes the hadronic activity in the event. We include the scalar sum of the transverse momenta of all reconstructed jets, the transverse momentum and pseudorapidity of the leading jet, the jet multiplicity and the leading-jet momentum fraction,
\begin{equation}
 \Big\{ H_T, \quad p_T^{j_1}, \quad \eta^{j_1}, \quad N_j, \quad p_T^{j_1}/H_T \Big\} .   
\end{equation}
These variables help distinguish the signal, which contains a high-multiplicity hadronic system originating from four $W$ bosons, from the background contributions in which the extra jets arise mainly from QCD radiation. We also consider a set of missing-transverse-momentum-related observables, 
\begin{equation}
    \Big\{E_T^{\rm miss}, \quad \phi(E_T^{\rm miss}), \quad M_{\rm eff}, \quad H_T/E_T^{\rm miss} \Big\} ,
\end{equation}
comprising the magnitude of the missing transverse momentum, its azimuthal direction, the effective mass defined by $M_{\rm eff}=H_T+E_T^{\rm miss}$ and the ratio of the hadronic activity to the missing transverse energy. We further include the minimum azimuthal separation between the missing transverse momentum and the jets, the azimuthal separations between the missing transverse momentum and the two leading jets, 
\begin{equation}
   \Big\{  \min_j\Delta\phi(E_T^{\rm miss},j) , \quad \Delta\phi(E_T^{\rm miss},j_{1,2}) \Big\} \,,
\end{equation}
and the transverse mass
\begin{equation}
  M_T(j_1,E_T^{\rm miss}) = \sqrt{2p_T^{j_1}E_T^{\rm miss} \left[ 1-\cos\Delta\phi(E_T^{\rm miss},j_1) \right]} .
\end{equation}
This missing-momentum-related set of variables is, however, not intended to exploit a genuine missing-energy signature of the signal. At a lepton collider, the initial transverse momentum is known, and any sizable $E_T^{\rm miss}$ in the signal therefore mainly reflects detector effects, jet-energy mismeasurements, out-of-cone radiation, or semi-leptonic decays inside the jets rather than an invisible particle in the hard process. The same is largely true for the dominant backgrounds considered here, which do not contain prompt invisible particles after imposing hadronic weak-boson decay constraints. The discriminating power of these observables is therefore expected to arise more from the pattern and orientation of the missing transverse momentum relative to the jets than from its absolute magnitude. The observable $M_T(j_1,E_T^{\rm miss})$ should therefore be interpreted as an event-balance observable rather than as a transverse-mass estimator associated with a genuine invisible particle.

\begin{table*}\setlength{\tabcolsep}{8pt}\renewcommand{\arraystretch}{1.5}
    \centering
    \resizebox{\linewidth}{!}{\begin{tabular}{c|cccc}
    Hyperparameters & $N_j \geq 6,~ p_T^j >25~\rm{GeV} $ & $N_j \geq 6,~ p_T^j >30~\rm{GeV} $ & $N_j \geq 8,~ p_T^j >20~\rm{GeV} $ & $N_j \geq 8,~ p_T^j >25~\rm{GeV} $ \\
    \hline 
    \texttt{n\_estimators} & 300 & 300 & 500 & 500 \\
    \texttt{max\_depth} & 4 & 5 & 5 & 3 \\
    \texttt{learning\_rate} & 0.1 & 0.05 & 0.1 & 0.05 \\
    \texttt{subsample} & 1.0 & 0.8 & 0.7 & 0.8 \\ \texttt{colsample\_bytree} & 0.7 & 0.7 & 1.0 & 0.8\\
     \texttt{gamma} & 0.2 & 0.1 & 0.1 & 0.1 \\
     \texttt{min\_child\_weight} & 2 & 1 & 3 & 1 \\
     \texttt{reg\_alpha} & 0.2 & 0.4 & 0.4 & 0.2 \\
     \texttt{early\_stopping\_rounds} & 20 & 20 & 20 & 20 \\
    \end{tabular}}
   \caption{Optimized hyperparameters for the BDT classifiers trained for four jet-multiplicity and jet-$p_T$ configurations and for a representative signal benchmark with $m_{\Delta^{\pm\pm}}=400~{\rm GeV}$. The classifiers are trained with a binary logistic objective and the AUC is used as the optimization metric.}
    \label{tab:hyperpara}
\end{table*}

Finally, we include observables associated with the two best reconstructed diboson systems and with the two spectator muons. For the two best reconstructed $W$ candidates, we use their invariant mass, angular separation and the difference between their reconstruction-quality variables,
\begin{equation}
  \Big\{ m_{W_1W_2}, \quad \Delta R(W_1,W_2), \quad \Delta\chi^2(W_1,W_2)\Big\} ,
\end{equation}
with $\Delta\chi^2(W_1,W_2) \equiv \chi^2_{W_2}-\chi^2_{W_1}$. When the two selected $W$ candidates originate from the same doubly charged scalar decay, the variable $m_{W_1W_2}$ provides an approximate proxy for $m_{\Delta^{\pm\pm}}$, up to combinatorial ambiguities and finite jet-energy resolution. Its distribution is therefore expected to shift with the signal benchmark mass, which further motivates training a separate classifier for each value of $m_{\Delta^{\pm\pm}}$. We also include the invariant mass, the transverse momentum and the pseudorapidity of the reconstructed three-$W$ system (for the $N_j\geq 6$ setup) and four-$W$ system (for the $N_j\geq 8$ setup),
\begin{equation}
  \Big\{ m_{3W}, \  p_T^{3W}, \  \eta^{3W} \Big\} \quad \text{or}\quad
  \Big\{ m_{4W}, \  p_T^{4W}, \  \eta^{4W} \Big\} ,
\end{equation}
which characterize the global kinematics of the heavy scalar-pair system reconstructed from its hadronic decay products, thereby providing information complementary to the pairwise $W$ observables. On the other hand, the two final-state muons provide additional information on the event recoil and scattering topology. We therefore include their transverse momenta and pseudorapidities, together with the transverse momentum of the dimuon system and the angular separations between the two muons,
\begin{equation}
  \Big\{ p_T^{\mu_{1,2}}, \  \eta^{\mu_{1,2}}, \  p_T^{\mu_1\mu_2}, \  \Delta R(\mu_1,\mu_2), \  \Delta\eta(\mu_1,\mu_2) \Big\} .
\end{equation}

Some of all the considered observables probe related aspects of the reconstructed event topology, and are therefore expected to be correlated. This redundancy is not problematic in a BDT analysis: it allows the classifier to exploit the same underlying physics through complementary kinematic handles, while the training procedure determines which variables or combinations of variables carry the most discriminating power.

For each signal benchmark mass and each jet-multiplicity/$p_T^j$ threshold setup, we train a separate BDT classifier. The signal and background samples are split into a training and a test dataset respectively corresponding to 70\% and 30\% of the available events, using stratified sampling to preserve the relative class composition. During training, background events from the different processes are weighted by their respective cross sections and by the benchmark luminosity ${\cal L}=1~{\rm ab}^{-1}$ before being combined into a single background sample. The signal and background samples are further weighted such that the two classes have equal total weight. This class-balanced training procedure preserves the physical relative composition of the background while removing the overall class imbalance with the signal, which improves the stability of the classifier in the presence of rare signal events. 

The 70\% training fraction is further divided into an internal training sample and a validation sample, with an $80\%/20\%$ split. The validation sample is used both to monitor the training and to implement early stopping, with the area under the receiver-operating-characteristic curve (AUC) serving as a validation metric~\cite{Hastie:2009itz}. In practice, the BDT hyperparameters are optimized by maximizing the AUC, which quantifies the overall ability of the classifier to rank signal events above background events. The training process is then stopped once the validation AUC no longer improves for a fixed number of boosting rounds. Following the validation strategy adopted in Ref.~\cite{Dehghani:2025xkd}, we monitor the AUC on statistically independent training, validation and test samples, and its stability across these samples provides a check that the classifier performance is not driven by overtraining. As an illustration, Table~\ref{tab:hyperpara} shows the optimized configurations obtained for the benchmark mass $m_{\Delta^{\pm\pm}}=400~{\rm GeV}$, for four choices of jet multiplicity and jet-$p_T$ threshold in $\mu^+\mu^+$ collisions at $\sqrt{s}=2~{\rm TeV}$. Finally, the independent test sample is kept aside and used only for performance evaluation. In this case, the event selection is obtained by applying a threshold on the BDT score, chosen by maximizing the expected significance computed from Eq.~\eqref{eq:significance} using the physical signal and background normalizations. 

\begin{figure*}
  \includegraphics[width=0.85\textwidth,trim={0 0 0 2cm},clip]{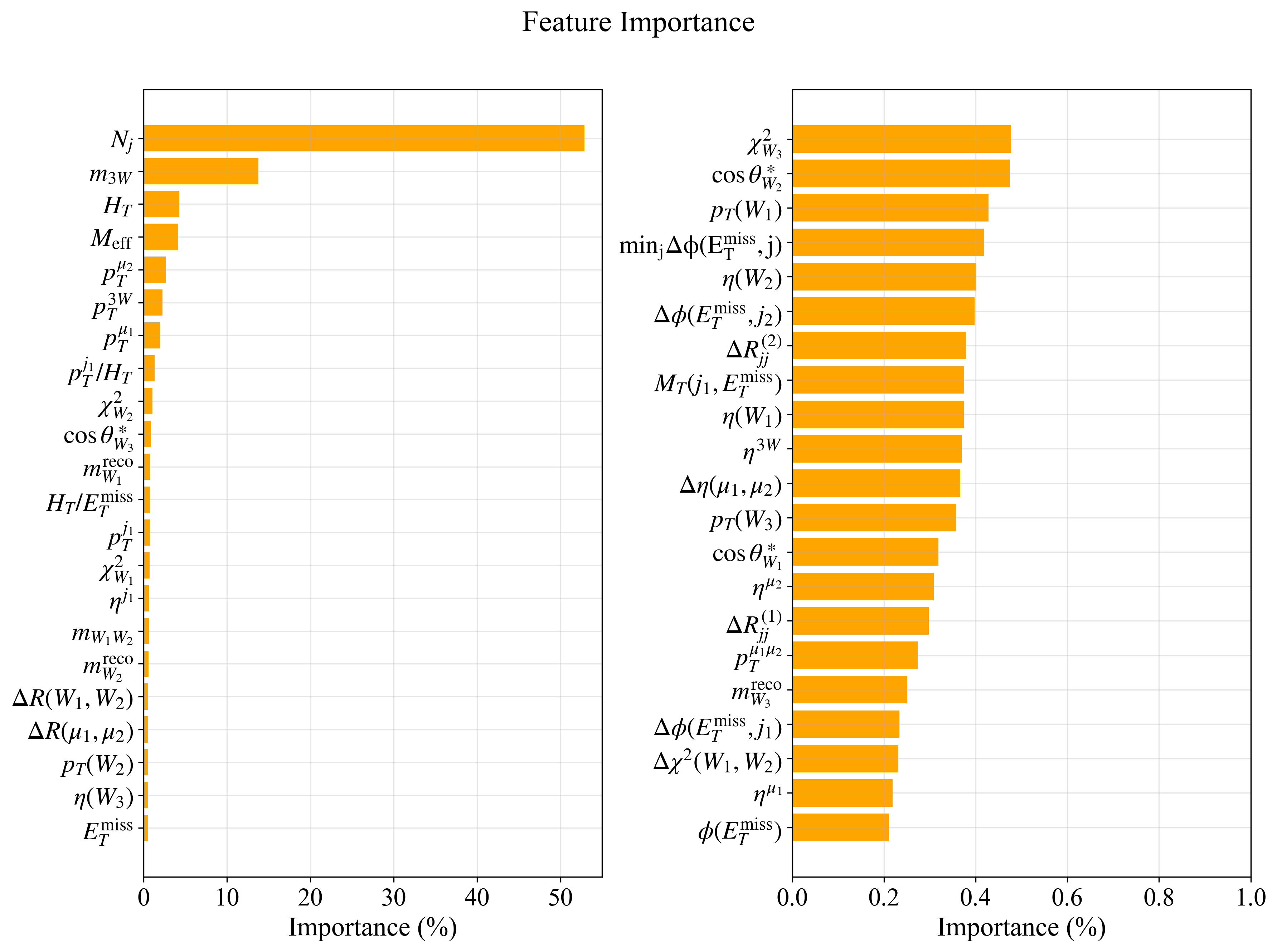}\\[.5cm]
  \includegraphics[width=0.85\textwidth,trim={0 0 0 2cm},clip]{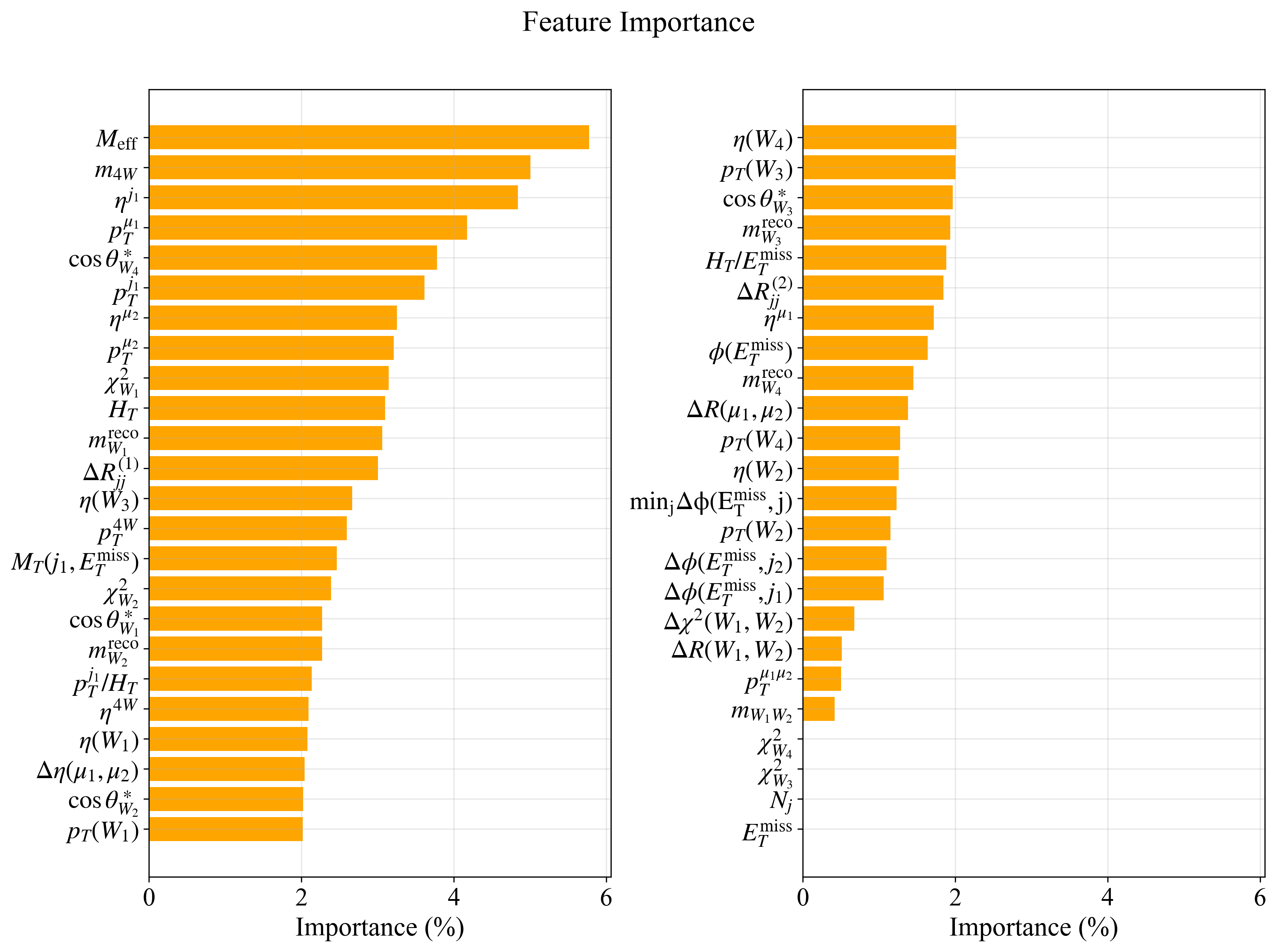}
  \caption{Gain-based feature importance of the input observables in the BDT classifiers for a representative signal benchmark with $m_{\Delta^{\pm\pm}}=400~{\rm GeV}$, in $\mu^+\mu^+$ collisions at $\sqrt{s}=2~{\rm TeV}$. The upper row corresponds to the $N_j\geq6,\ p_T^j>25~{\rm GeV}$ setup while the lower row corresponds to the $N_j\geq8,\ p_T^j>20~{\rm GeV}$ setup, and in each row the ranked list of variables is split into two panels for readability. The importances are computed using the gain metric implemented in \textsc{XGBoost} and normalized to the total gain of the corresponding classifier.}
\label{fig:feature_significance_W_BDT}
\end{figure*}

We first examine which input observables drive the BDT discrimination. For this purpose, we use the gain-based feature-importance metric implemented in \textsc{XGBoost}, which measures the average improvement of the splitting criterion obtained when a given variable is used in the decision trees. This should therefore be interpreted as a qualitative diagnostic of the trained classifier rather than as a direct measure of the expected significance associated with an individual observable. Figure~\ref{fig:feature_significance_W_BDT} shows the resulting feature rankings for the same representative benchmark as above, defined by $m_{\Delta^{\pm\pm}}=400~{\rm GeV}$, and for two illustrative analysis setups. The upper row corresponds to the relaxed selection $N_j\geq6$ with \mbox{$p_T^j>25~{\rm GeV}$}, while the lower row corresponds to the more resolved selection $N_j\geq8$ with $p_T^j>20~{\rm GeV}$. The two cases are found to exhibit rather different patterns, which reflect the different background compositions. However, since the figure corresponds to a single representative benchmark, the precise ordering of the variables should not be interpreted as universal across the full mass scan. In particular, observables tied directly to the reconstructed multi-$W$ invariant masses may change in relative importance as $m_{\Delta^{\pm\pm}}$ is varied.

For the $N_j\geq6$ setup, the ranking is strongly hierarchical and is dominated by the jet multiplicity. This originates from the fact that once the $N_j\geq8$ multiplicity requirement is relaxed, the selected background sample contains many more events in which the additional jets arise from QCD radiation rather than from a genuine four-$W$ topology. The BDT therefore recovers a substantial part of the discrimination lost by loosening the cut through the residual information contained in $N_j$. In this sense, the classifier effectively implements a softer version of the $N_j\geq8$ multiplicity requirement used in the cut-based analysis, rather than treating all events passing $N_j\geq6$ on an equal footing. 

The following most relevant variables, such as the reconstructed multi-$W$ invariant mass $m_{3W}$, the hadronic activity $H_T$, the effective mass $M_{\rm eff}$ and the transverse momentum of the reconstructed multi-$W$ system $p_T^{3W}$, probe the overall hard scale of the event. Their appearance in the ranking is consistent with the signal being characterized by the decay of a heavy scalar pair into several massive weak bosons, whereas the dominant backgrounds populate the high-multiplicity region mainly through radiation. The relatively large importance of the spectator-muon transverse momenta is also meaningful: these clean leptonic observables retain information on the scattering kinematics and on the recoil against the heavy hadronic system, and therefore become particularly useful when the hadronic topology is only partially reconstructed. 

The situation changes once the more resolved $N_j\geq8$ requirement is imposed. In this case, the jet multiplicity itself carries little additional information since both signal and surviving background events already satisfy a high-multiplicity condition. The feature importance is instead distributed over a broader set of observables, with $M_{\rm eff}$ providing the leading contribution, followed by variables such as $m_{4W}$, $\eta^{j_1}$, $p_T^{\mu_1}$ and $\cos\theta^*_{W_2}$. 

The prominence of $M_{\rm eff}$ and $m_{4W}$ reflects the importance of the overall hard scale and of the reconstructed heavy multi-boson system. The sizable role of the leading jet pseudorapidity $\eta^{j_1}$ indicates that the classifier also exploits the geometric structure of the hadronic activity: jets originating from the decay of a heavy multi-$W$ system need not populate the same pseudorapidity domain as jets arising from radiation-dominated background topologies. Similarly to the $N_j\geq6$ setup, the importance of the variable $p_T^{\mu_1}$ shows that the spectator muons retain useful information on the underlying same-sign scattering kinematics and on the recoil against the heavy hadronic system. In addition, the appearance of $\cos\theta^*_{W_2}$ confirms that the BDT is not only using global energy-scale variables, but also information on the internal decay geometry of the reconstructed $W$ candidates. This sensitivity should be understood at the reconstruction level: the helicity-angle variables help distinguish genuine hadronic $W$ candidates from accidental dijet combinations, without requiring an explicit interpretation in terms of a charge-tagged quark direction. Finally, the relatively modest importance of $\Delta\chi^2(W_1,W_2)$ is also not unexpected. This observable is derived from the individual reconstruction-quality variables and is therefore partly redundant with the $\chi^2_{W_1}$ and $\chi^2_{W_2}$ variables. The BDT can thus exploit the quality of the leading $W$ candidates directly, without necessarily assigning a large gain to their difference. 

\begin{figure}
  \centering
  \includegraphics[width=0.99\columnwidth]{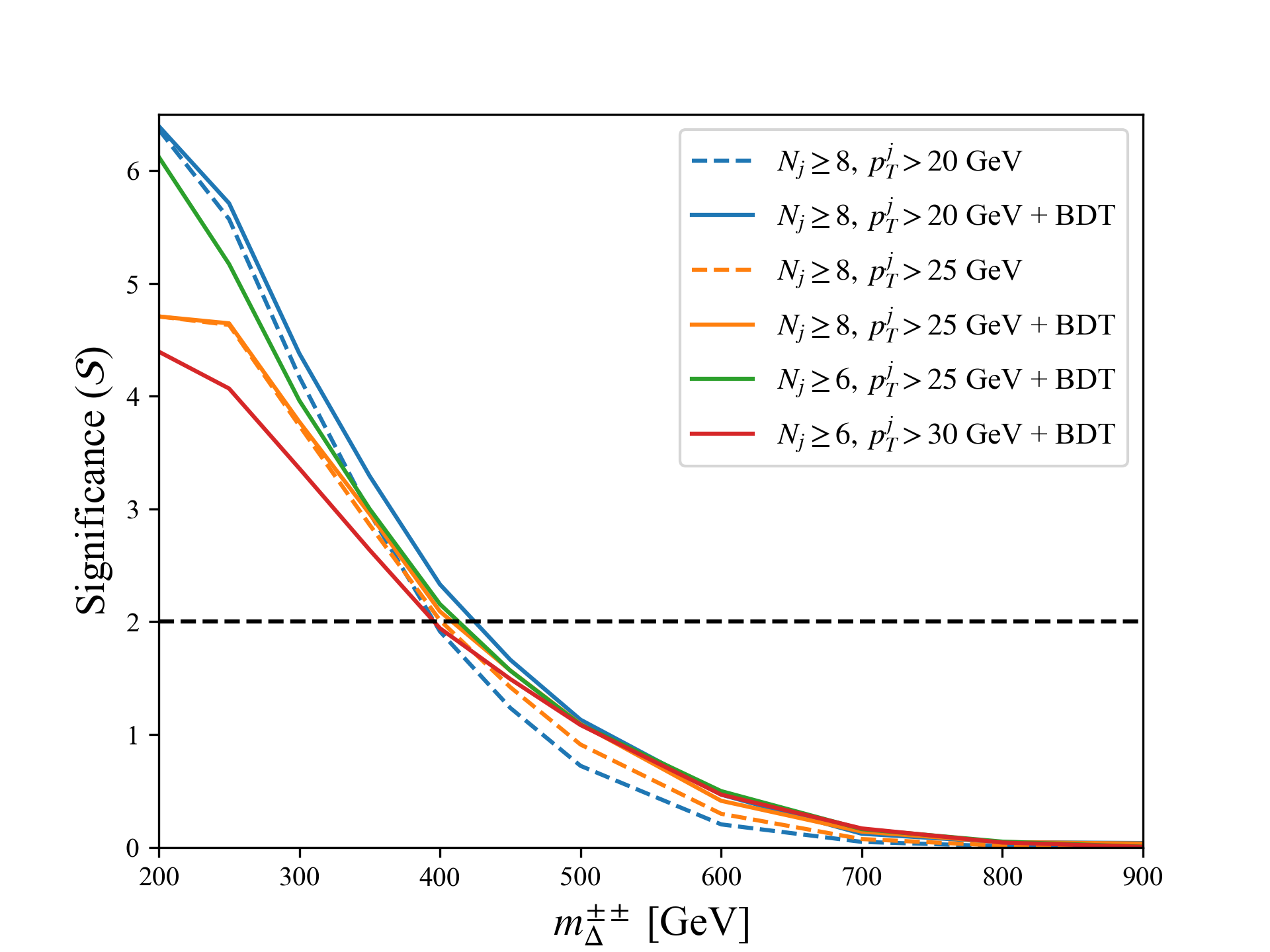}
  \caption{Expected signal significance as a function of the doubly charged scalar mass in $\mu^+\mu^+$ collisions at $\sqrt{s}=2~{\rm TeV}$ and for ${\cal L}=1~{\rm ab}^{-1}$. The solid curves show the BDT-based results obtained after imposing jet-multiplicity and jet-$p_T$ preselection requirements: \mbox{$N_j\geq8$}, $p_T^j>20~{\rm GeV}$ (blue), $N_j\geq8,\ p_T^j>25~{\rm GeV}$ (orange), $N_j\geq6,\ p_T^j>25~{\rm GeV}$ (green) and $N_j\geq6,\ p_T^j>30~{\rm GeV}$ (red). The dashed curves show the corresponding $2tW$ cut-based results for the two $N_j\geq8$ selections.}
    \label{fig:signi_mdel}
\end{figure}

It is also worth noting that the missing-transverse-momentum variables do not dominate the ranking, which is consistent with the fact that the signal does not contain a genuine invisible particle in the hard process. When such variables enter the classifier, their role is instead connected to event balance and to the orientation of the missing transverse momentum relative to the reconstructed jets rather than to its absolute magnitude, as anticipated. In addition, this demonstrates that the prominence of $M_{\rm eff}$ in the ranking stems mostly from its dominant $H_T$ component, since $E_T^{\rm miss}$ alone ranks much lower. 

Nevertheless, feature importances in a multivariate classifier should be interpreted with some care. Correlations between input variables can redistribute the gain among observables that encode related physics, while tree-based importance measures may also favor observables with many possible split points. The feature ranking is therefore used as a qualitative diagnostic of the trained classifier rather than as a model-independent measure of the intrinsic discriminating power of each variable~\cite{Hastie:2009itz, Cornell:2021gut}.

We now assess the impact of the multivariate analysis on the expected sensitivity. Figure~\ref{fig:signi_mdel} shows the signal significance as a function of $m_{\Delta^{\pm\pm}}$ for several jet-multiplicity and jet-$p_T$ threshold choices. The solid curves correspond to the BDT-based analysis, while the dashed curves show the corresponding $2tW$ cut-based reference results for the two $N_j\geq8$ selections. In all cases the significance decreases with increasing $m_{\Delta^{\pm\pm}}$ values, reflecting the rapid decrease of the production cross section and the corresponding reduction of the signal yield at large masses.

The BDT analysis improves the sensitivity relative to the corresponding cut-based strategy for all jet-multiplicity and jet-$p_T$ threshold choices considered, although only the two $N_j\geq8$ cut-based reference curves are shown in Figure~\ref{fig:signi_mdel} to avoid overcrowding the figure. The improvement is moderate rather than dramatic, and it follows from several competing effects. At low masses, the curves cluster within about two units of significance, within ${\cal S}\simeq 4.5-6.5$. For such light doubly charged scalars the production rate is still large enough that all selections retain a sizable signal yield. The sensitivity is therefore relatively insensitive to the precise choice of the jet multiplicity requirement, the jet-$p_T$ threshold or even the BDT-score requirement. In this regime, the analysis is already efficient and the multivariate classifier mainly refines a separation that is not yet limited by the smallness of the signal sample. 

The separation between the different analysis choices is largest in the intermediate mass range, around $m_{\Delta^{\pm\pm}}\simeq 350-500~{\rm GeV}$, where the BDT shifts the ${\cal S}=2$ reach to about $425-430~{\rm GeV}$ for the best-performing setup. This is the regime where the analysis optimization has also the greatest phenomenological impact: the signal yield is no longer so large that all selections perform similarly, but it is still large enough for improved kinematic discrimination to translate into a visible gain in significance. 

At higher masses, the analysis becomes instead rate-limited. The signal cross section decreases by more than four orders of magnitude between $m_{\Delta^{\pm\pm}}=200~{\rm GeV}$ and $900~{\rm GeV}$ at $\sqrt{s}=2~{\rm TeV}$, as shown in Figure~\ref{fig:sigma_mdel}, leading to a rapid reduction of the expected signal yield. In the low-background regime relevant after the multivariate selection, the significance scales approximately as $\sqrt{\sigma_{\rm sig}}$ when the surviving signal dominates the denominator in Eq.~\eqref{eq:significance}, and even more steeply when the residual background becomes important. As a result, improvements in event selection can only provide limited gains, and overcoming this suppression would require a substantial increase in integrated luminosity or center-of-mass energy. This can be made more quantitative by a simple luminosity-scaling estimate~\cite{Araz:2019otb}. Keeping the same selection, both the signal and background yields scale linearly with the integrated luminosity, so that the significance scales as ${\cal S}\propto\sqrt{\cal L}$. The luminosity required to reach a target significance ${\cal S}_\star$ is therefore given by
\begin{equation}
  {\cal L}_\star = 1~{\rm ab}^{-1} \bigg[ \frac{{\cal S}_\star}{{\cal S}(1~{\rm ab}^{-1})} \bigg]^2 .
\end{equation}
For the best-performing BDT setup, this estimate shows that a $400~{\rm GeV}$ doubly charged scalar would require about $6~{\rm ab}^{-1}$ to reach the ${\cal S}=5$ discovery regime, which is demanding but still within the range of a high-luminosity future-collider discussion. In contrast, for $m_{\Delta^{\pm\pm}}\simeq700~{\rm GeV}$, the luminosity required to reach even ${\cal S}=2$ is of order $10^3~{\rm ab}^{-1}$, far beyond any realistic expectation. Whereas it is plausible that a higher-energy same-sign muon collider configuration would alleviate the near-threshold suppression and extend the reach to larger scalar masses, such a study would require dedicated center-of-mass and luminosity assumptions beyond the $\mu$TRISTAN configuration adopted in this work~\cite{Hamada:2022mua}.

Among the considered BDT configurations, the \mbox{$N_j\geq8$}, $p_T^j>20~{\rm GeV}$ selection gives the largest significance over most of the mass range. This illustrates the advantage of retaining the high-multiplicity requirement, which preserves the resolved four-$W$ topology and keeps the background under control, while lowering the jet-$p_T$ threshold recovers signal events containing softer jets from hadronic $W$ decays. This effect is particularly relevant at low and intermediate masses, where the $W$ bosons from the scalar decay are not yet highly boosted and one of the two jets from a hadronic $W$ decay can more easily fall close to the analysis threshold. Accordingly, the $N_j\geq8,\ p_T^j>25~{\rm GeV}$ setup is slightly less sensitive, but remains close to the best curve. 

The relaxed $N_j\geq6$ selections provide an instructive comparison. As discussed in Section~\ref{subsec:Cutbased_analysis}, going from the baseline $N_j\geq8,\ p_T^j>25~{\rm GeV}$ selection to a relaxed $N_j\geq6,\ p_T^j>30~{\rm GeV}$ selection raises the background from about $8.2~{\rm ab}$ to about $0.68~{\rm fb}$, \textit{i.e.}\ by about two orders of magnitude. The BDT can exploit the residual jet-multiplicity information and the additional kinematic variables discussed above, but it does not fully compensate for the loss of purity. The $N_j\geq6$ sensitivity curves therefore remain below the ones related to the resolved $N_j\geq8$ configurations, the effect being particularly visible for the $p_T^j>30~{\rm GeV}$ threshold choice where the harder jet requirement further reduces the signal acceptance and leads to the weakest sensitivity among all BDT setups shown.

It is finally useful to place this reach in the context of existing LHC searches. As discussed in Section~\ref{sec:model}, the current direct LHC sensitivity to doubly charged scalars decaying dominantly into same-sign $W$ boson pairs reaches the few-hundred-GeV range, with the strongest relevant bounds lying around $m_{\Delta^{\pm\pm}}\simeq 350~{\rm GeV}$~\cite{ATLAS:2018ceg, ATLAS:2021jol}. The best-performing $\mu$TRISTAN BDT setup reaches a reference sensitivity level of ${\cal S}=2$ for $m_{\Delta^{\pm\pm}}\simeq 425-430~{\rm GeV}$, corresponding to an indicative extension of the direct diboson coverage by roughly $70-80~{\rm GeV}$. This comparison again shows that same-sign muon collisions provide an independent and complementary probe of the diboson-dominated Type-II seesaw parameter space, precisely in the regime where conventional leptonic searches for the production and decay of $\Delta^{\pm\pm}$ scalars lose sensitivity.

\section{Summary and Conclusions} \label{sec:summary}

Doubly charged scalars $\Delta^{\pm\pm}$ are a distinctive prediction of several extensions of the Standard Model. In the Type-II seesaw framework, they arise as part of an ${\rm SU}(2)_L$ scalar triplet whose vacuum expectation value $v_t$ controls both the neutrino-mass generation mechanism and the collider phenomenology of the triplet states. For small triplet vevs, the doubly charged scalar predominantly decays into same-sign charged leptons, leading to the conventional leptonic and lepton-flavor-violating search channels. In contrast, for larger values of the triplet vev, the diboson decay mode $\Delta^{\pm\pm}\to W^\pm W^\pm$ becomes dominant, and the sensitivity of the conventional leptonic searches is correspondingly reduced. 

In this work, we have investigated this large-$v_t$ diboson-dominated regime at a same-sign $\mu^+\mu^+$ collider, motivated by the $\mu$TRISTAN proposal. We have focused on a benchmark scenario with a degenerate triplet spectrum and a triplet vev $v_t=1~{\rm GeV}$, for $\sqrt{s}=2~{\rm TeV}$ and an integrated luminosity of $1~{\rm ab}^{-1}$. The signal topology consists of two spectator muons and four $W$ bosons, $\mu^+\mu^+ \to \mu^+\mu^+ \Delta^{++}\Delta^{--} \to \mu^+\mu^+ W^+W^+W^-W^-$, with the $W$ bosons taken to decay hadronically. This leads to a challenging but characteristic $\mu^+\mu^+ + 8j$ final state, which we examine through complementary cut-based and multivariate strategies.

We have first developed a cut-based analysis exploiting the high jet multiplicity and the possibility of reconstructing several hadronic $W$ candidates. The dominant backgrounds arise from same-sign muon scattering in association with QCD radiation and weak-boson production, and a high jet-multiplicity requirement is essential to suppress them, as they contain fewer genuine hadronic weak bosons. We have reconstructed the hadronic $W$ candidates through a global $\chi^2$ minimization procedure over all possible dijet pairings in the events, and defined several nested signal regions based on the quality of the reconstructed $W$ candidates. Such a cut-based strategy turns out to reduce the background to the ab level and provide sensitivity to doubly charged scalars in the few-hundred-GeV mass range. 

We have then improved the analysis using a multivariate strategy based on boosted decision trees. The BDTs jointly exploit a broad set of observables, including reconstructed-$W$ kinematics and quality variables, global hadronic activity, missing-transverse-momentum-related event-balance variables, reconstructed multi-$W$ quantities, spectator-muon kinematics and angular observables. The best-performing multivariate configuration is obtained for a resolved preselection requiring $N_j\geq8$ and $p_T^j>20~{\rm GeV}$. In this case, we obtain a $2\sigma$ sensitivity for masses up to $m_{\Delta^{\pm\pm}}\simeq425-430~{\rm GeV}$, thereby extending the reach of the corresponding cut-based strategy by a few tens of GeV. The improvement is most relevant in the intermediate-mass region, where the signal yield is no longer large enough for cut-based selections to perform well, but is still sufficient for improved kinematic discrimination to translate into a visible gain in significance. At larger masses, the sensitivity becomes rapidly rate-limited, and our results indicate that substantially extending the high-mass reach would require either a much larger luminosity or a higher center-of-mass energy.

Finally, it is useful to compare these results with existing LHC searches in the diboson-dominated regime which probe doubly charged scalars up to the few-hundred-GeV range, with the strongest relevant bounds around $m_{\Delta^{\pm\pm}}\simeq 350~{\rm GeV}$. The $\mu$TRISTAN strategy studied here therefore provides an indicative extension of the direct diboson coverage by roughly $70-80~{\rm GeV}$, which demonstrates that same-sign muon collisions offer an independent and complementary probe of the Type-II seesaw parameter space precisely in the region where leptonic searches for $\Delta^{\pm\pm}$ lose sensitivity. 

\acknowledgments

The authors are grateful to Songshaptak De for useful discussion regarding the BDT analysis. The work of MF has been partly supported by NSERC through grant number SAP105354. 

\bibliographystyle{utcaps_mod}
\bibliography{Muon}
\end{document}